\documentclass[a4paper,fleqn,usenatbib]{mnras}

\usepackage{graphicx}
\usepackage{bm}
\usepackage{hyperref}
\usepackage{amsmath}
\usepackage{amssymb}
\usepackage{booktabs}
\usepackage{siunitx}
\usepackage{xcolor}
\usepackage{float}
\usepackage{microtype}
\hypersetup{colorlinks=true,citecolor=blue,linkcolor=blue,urlcolor=blue}
\DeclareRobustCommand{\VAN}[3]{#2}
\let\VANthebibliography\thebibliography
\def\thebibliography{\DeclareRobustCommand{\VAN}[3]{##3}%
  \VANthebibliography}

\newcommand{\NEW}[1]{#1}

\title[Dynamical--Photometric Phase Space of Spiral Galaxies]{%
  A Dynamical--Photometric Phase Space for Spiral Galaxies:
  Probing the Local Coupling Between Light and Gravity}

\author[A.~Sanyal \& F.~Rahaman]{%
  Aritra Sanyal$^{1}$\thanks{E-mail: aritrasanyal1@gmail.com}
  and
  Farook Rahaman$^{1}$\thanks{E-mail:
    rahaman@associates.iucaa.in}
  \\
  $^{1}$Department of Mathematics, Jadavpur University,
  Kolkata~700\,032, India}

\begin{document}
\maketitle

\begin{abstract}
The interplay between luminous matter distribution and the
local gravitational field within disc galaxies encodes
physical information beyond that captured by global scaling
relations. We introduce a
\emph{dynamical--photometric phase space} defined by the
kinematic variable $X(R)=V(R)/R$ and the photometric
variable $Y(R)=\mathrm{d}\ln I/\mathrm{d}\ln R$, placing
the local gravitational scale and the logarithmic surface
brightness gradient into direct pointwise correspondence at
each galactocentric radius $R$. The quantity
$X=V/R=\omega$ represents the angular frequency of circular
motion and acts as a probe of the local mean mass density,
while $Y$ measures the radial steepness of the stellar light
distribution. The baryon-dominated inner disc is
characterized by large negative $Y$, whereas the
dark-matter-dominated outer region approaches
$Y\rightarrow0$. This two-regime behaviour is described by
the smooth sigmoid relation
$Y=[a\ln X+b]/(1+\exp[k(X-X_{\rm trans})])$, which reduces
to the logarithmic coupling $Y=a\ln X+b$ in the baryonic
zone. We apply this framework to 136 late-type galaxies
from the SPARC database, spanning inclinations
$20^{\circ}$--$89^{\circ}$, distances $1$--$130$\,Mpc,
and five decades in stellar mass. The median coefficient
of determination is $R^{2}=0.930$. Statistical validation
includes eight independent tests together with 5-fold
cross-validation. The transition parameter
$X_{\rm trans}$ identifies the onset of dark-matter
dominance, corresponding to a median transition radius
$R_{\rm trans}=5.40$\,kpc across the sample.
\end{abstract}


\section{Introduction}
\label{sec:intro}

The coupling between the visible baryonic component of a
galaxy and its gravitational field represents one of the
most actively debated problems in extragalactic astrophysics.
Observations of rotation curves have established beyond
doubt that the circular velocities of spiral galaxies
cannot be explained by the contribution of stars and gas
alone \citep{rubin1978,rubin1980,bosma1981,sofue2001}.
The implied presence of an extended, dynamically dominant
dark-matter halo \citep{ostriker1973,einasto1974,nfw1997}
or, alternatively, a modification of Newtonian dynamics on
galactic scales \citep{milgrom1983,famaey2012,mcgaugh2004},
has motivated decades of observational and theoretical work.

A central tool in this endeavour has been the construction
of empirical scaling relations that connect observable
photometric and kinematic properties of galaxies. The
Tully--Fisher relation \citep{tf1977}, which links the
total baryonic mass to the asymptotic rotation velocity,
has been refined into the baryonic Tully--Fisher relation
\citep{mcgaugh2000,lelli2016a,lelli2019}, now recognised
as one of the tightest scaling relations in galaxy physics
with intrinsic scatter below 0.1\,dex. The central
surface density relation \citep{lelli2016b_cdr} establishes
a tight link between the central dynamical surface density
and the central stellar surface density across galaxy
types. Most compellingly, the radial acceleration relation
(RAR; \citealt{mcgaugh2016,lelli2017a}) reveals a
point-by-point correspondence between the total centripetal
acceleration $g_{\rm tot}=V^{2}/R$ and the acceleration
expected from baryons alone $g_{\rm bar}$, confirmed
across $\sim$2700 data points from 153 SPARC galaxies
and subsequently extended to dwarf spheroidals
\citep{lelli2017b}, early-type galaxies \citep{lelli2017a},
and galaxy clusters \citep{chan2020}. Together these
relations suggest a profound and potentially universal
coupling between baryonic mass distribution and the
gravitational field.

Yet all of these relations are fundamentally \emph{global}
in character: they compare integrated or averaged
quantities --- total luminosity, asymptotic velocity,
enclosed mass --- and cannot reveal how the baryon--gravity
coupling varies \emph{locally} across the disc as a
function of galactocentric radius. In particular, none of
them directly probes the radial gradient of the stellar
light distribution and its connection to the local
kinematic state at each individual radius. The radial
surface brightness profile $I(R)$ and its logarithmic
derivative $\mathrm{d}\ln I/\mathrm{d}\ln R$ contain rich
information about the radial organisation of the stellar
disc --- the scale length, the bulge--disc transition,
bar and ring features, the approach to the flat outer
profile --- that is washed out when the profile is
integrated over radius or averaged across galaxies.

In this paper we introduce a \emph{dynamical--photometric
phase space} that bridges this gap. By constructing the
phase space $(X,Y)$ where $X(R)=V(R)/R$ is the local
angular frequency and $Y(R)=\mathrm{d}\ln I/\mathrm{d}\ln R$
is the logarithmic surface brightness gradient, we obtain
a direct, local, pointwise comparison between kinematics
and photometric structure at every radius simultaneously.
This framework carries three distinct advantages over
existing scaling relations. First, it does not require
knowledge of the stellar mass-to-light ratio: $Y$ is a
purely photometric quantity and $X$ is a purely kinematic
quantity, so no mass conversion is needed. Second, it
resolves the baryon--gravity coupling as a function of
radius, distinguishing the baryon-dominated inner disc
from the dark-matter-dominated outer disc. Third, the
transition between these two regimes is parametrised
directly by the sigmoid model, yielding the dark-matter
onset radius $R_{\rm trans}$ as a directly measurable
quantity for each individual galaxy.

The logarithmic coupling $Y=a\ln X+b$ that describes the
inner baryon zone has a theoretical basis: it emerges
approximately from the combination of an exponential
stellar disc \citep{freeman1970} with a self-gravitating
Freeman-disc rotation curve \citep{binney2008}, as we
show in Section~\ref{sec:methods}. This theoretical
motivation --- albeit approximate --- distinguishes the
logarithmic form from an arbitrary empirical choice and
gives the slope $a$ and transition scale $X_{\rm trans}$
physically interpretable meanings.

We apply the framework to 136 late-type galaxies from the
SPARC database \citep{lelli2016b}. We present detailed
results for ten representative galaxies spanning
$i=20^\circ$--$89^\circ$ and report general trends across
the full 136-galaxy sample. Section~\ref{sec:data}
describes the data and error propagation.
Section~\ref{sec:methods} derives the theoretical basis
of the phase space and describes the sigmoid model and
statistical validation. Section~\ref{sec:results} presents
all results. Section~\ref{sec:disc} discusses the physical
interpretation. Section~\ref{sec:conc} summarises our
conclusions.

\section{Data}
\label{sec:data}

\subsection{The SPARC database}
\label{sec:sparc}

All rotation curves and stellar surface brightness profiles
used in this work are drawn from the SPARC (Spitzer
Photometry and Accurate Rotation Curves) database
\citep{lelli2016b}. SPARC provides homogeneously reduced
kinematic and photometric data for late-type galaxies
(spirals and irregulars) compiled from the literature and
supplemented with Spitzer Space Telescope IRAC Channel~1
($3.6\,\mu$m) imaging, spanning five decades in stellar
mass ($10^{7}$--$10^{11.5}\,M_\odot$), three decades in
stellar surface brightness, and the full range of late
Hubble types from Sa to Im.

Rotation curves are compiled from interferometric
H\,{\sc i} observations and, for the inner regions,
H$\alpha$ long-slit spectroscopy and Fabry--P\'erot
interferometry; the original data references are tabulated
in \citet{lelli2016b}. Circular velocities are
inclination-corrected and represent the total (dark plus
baryonic) gravitational potential. Photometric profiles
are derived from Spitzer IRAC imaging at $3.6\,\mu$m,
which minimises contributions from hot dust, ionised gas
emission, and young stellar populations
\citep{meidt2014,querejeta2015}, and whose mass-to-light
ratio is well constrained and nearly independent of
stellar colour \citep{mcgaugh2014,schombert2019}. The
$3.6\,\mu$m band therefore provides the most reliable
available tracer of the stellar surface mass density
$\Sigma_\star(R)$ in nearby disc galaxies. We use a
sample of 136 SPARC galaxies that have sufficient overlap
between the kinematic and photometric radial grids and
whose sigmoid fit (Section~\ref{sec:sigmoid}) achieves
$R^{2}>0.3$. These 136 galaxies are the basis of all
results in this paper.

\subsection{Rotation curves and photometric profiles}
\label{sec:profiles}

\begin{figure*}
  \centering
  \includegraphics[width=\textwidth]{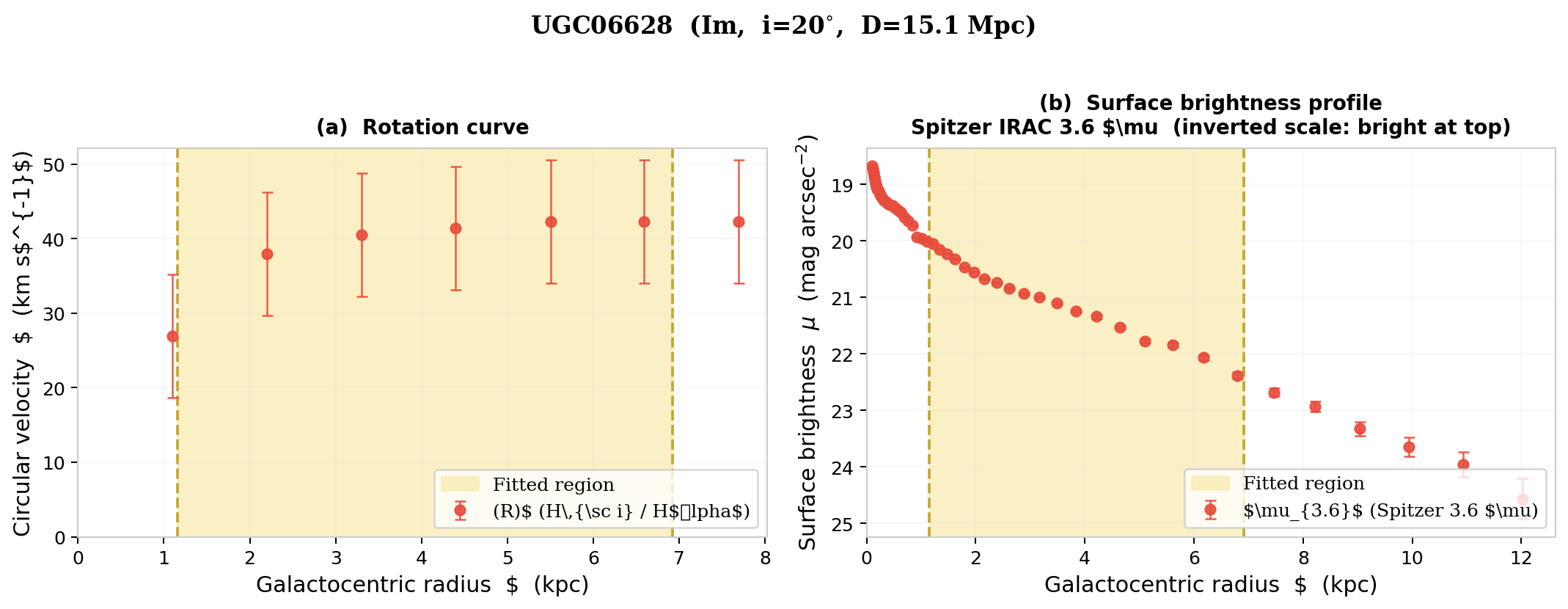}
  \caption{\textbf{UGC~06628} (Im, $i=20^\circ$, $D=15.1$\,Mpc).
    \textit{Left}: Circular velocity $V(R)$ versus
    galactocentric radius $R$ with $1\sigma$ error bars.
    \textit{Right}: Spitzer $3.6\,\mu$m surface brightness
    profile $\mu(R)$ on an inverted magnitude scale
    (brighter at top). Gold shaded bands mark the radial
    range used for phase-space fitting.}
  \label{fig:UGC06628_prof}
\end{figure*}

\begin{figure*}
  \centering
  \includegraphics[width=\textwidth]{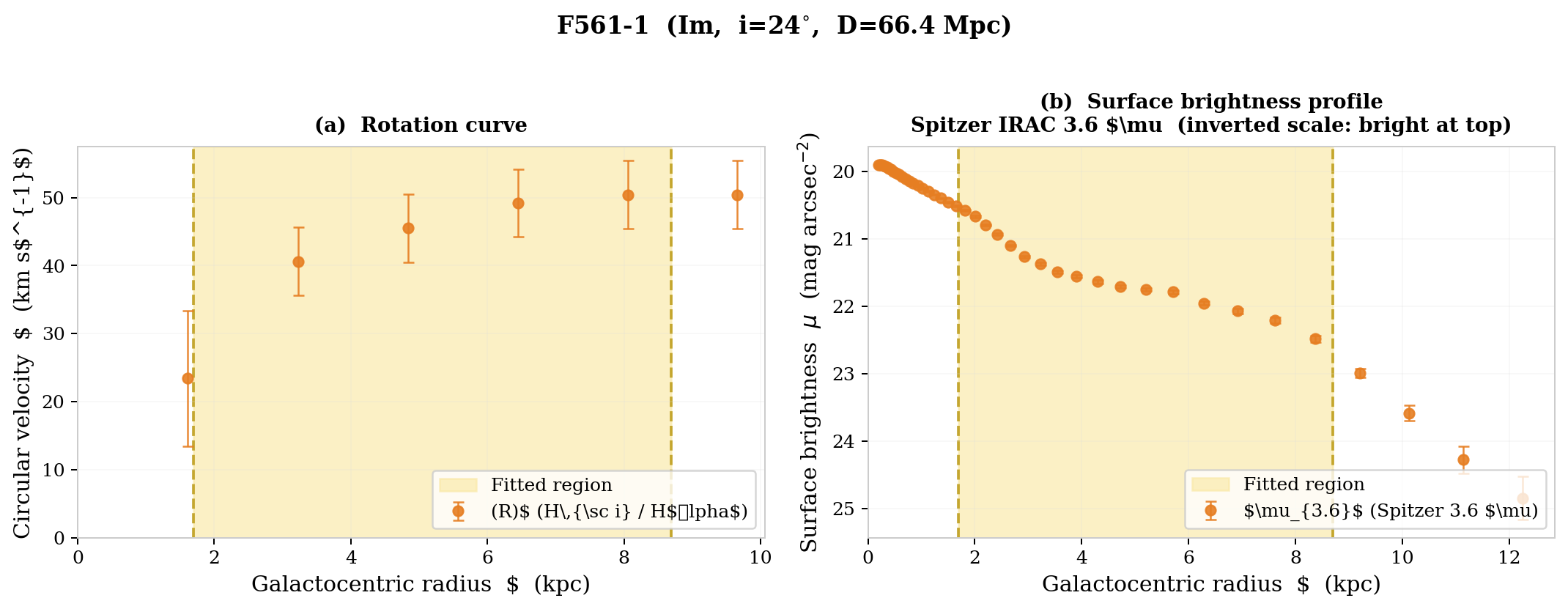}
  \caption{\textbf{F561-1} (Im, $i=24^\circ$, $D=66.4$\,Mpc).
    Panels as in Figure~\ref{fig:UGC06628_prof}.
    The most distant galaxy in the sample; the
    two-regime phase-space structure is cleanly
    detected at 66.4\,Mpc.}
  \label{fig:F5611_prof}
\end{figure*}

\begin{figure*}
  \centering
  \includegraphics[width=\textwidth]{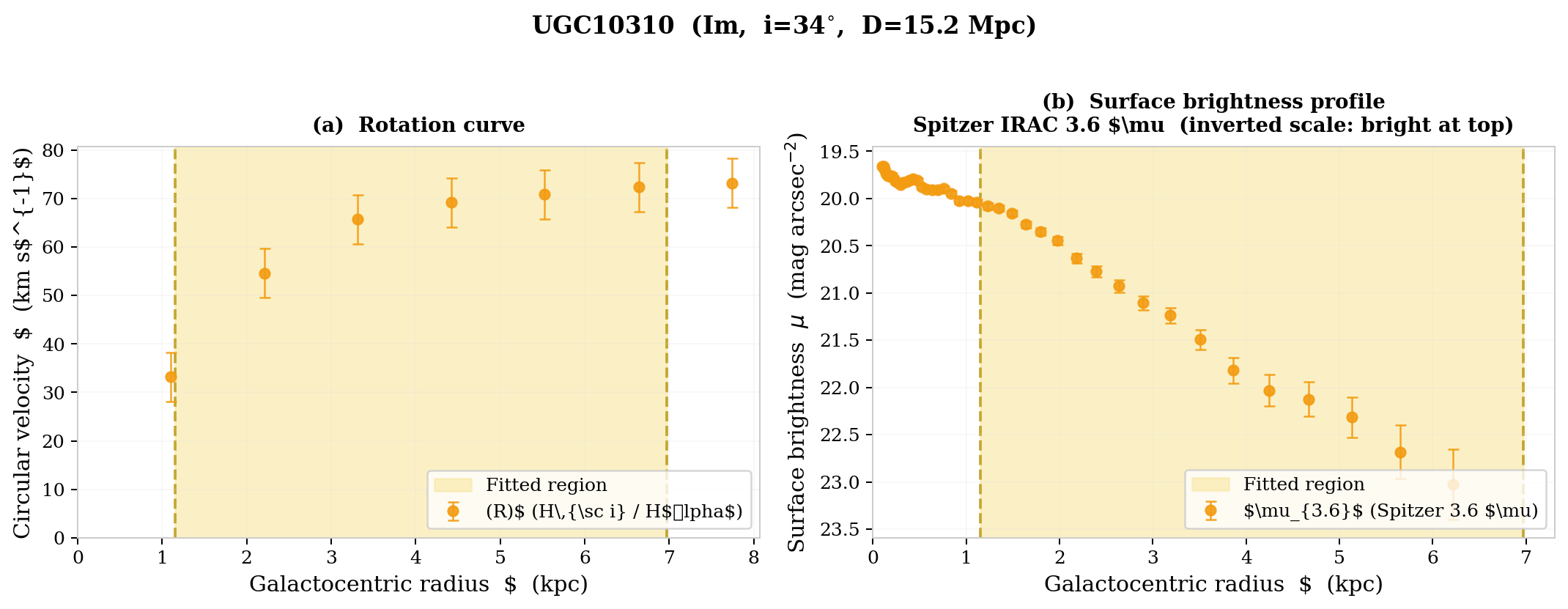}
  \caption{\textbf{UGC~10310} (Im, $i=34^\circ$, $D=15.2$\,Mpc).
    Panels as in Figure~\ref{fig:UGC06628_prof}.
    A low-mass dwarf irregular with a slowly rising
    rotation curve and a smooth exponential
    surface brightness profile.}
  \label{fig:UGC10310_prof}
\end{figure*}

\begin{figure*}
  \centering
  \includegraphics[width=\textwidth]{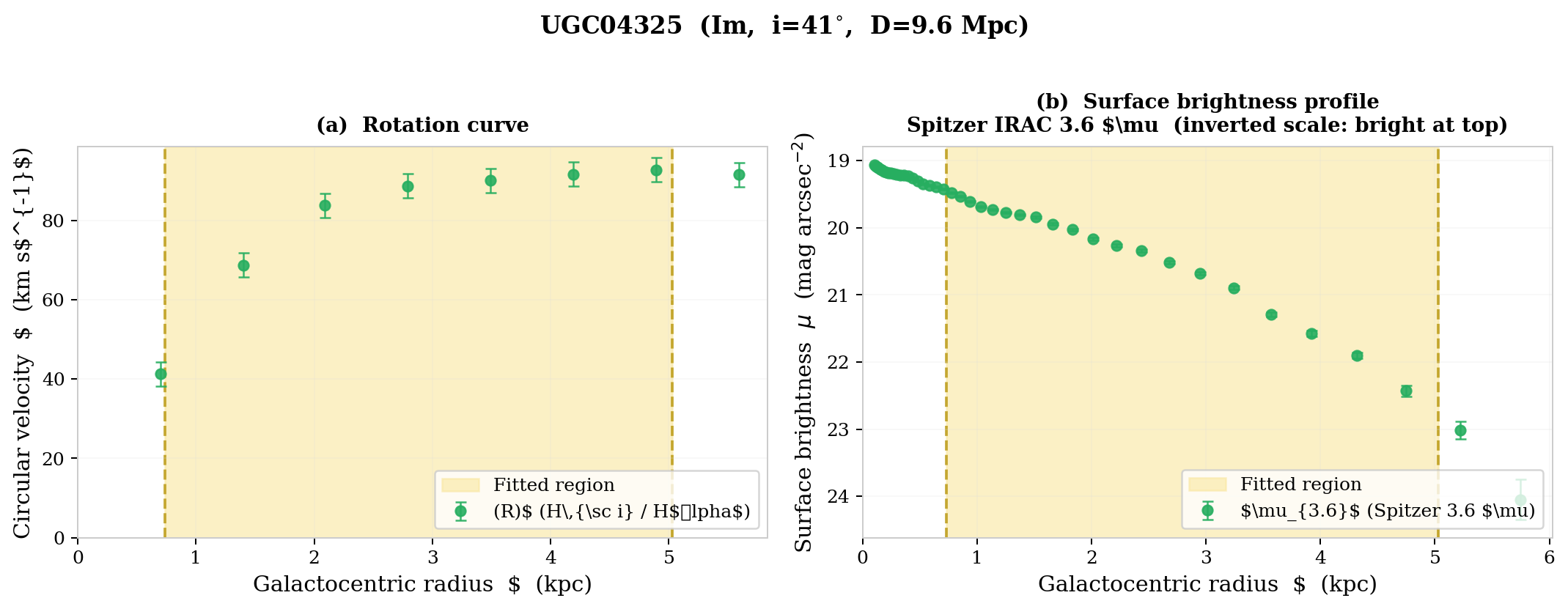}
  \caption{\textbf{UGC~04325} (Im, $i=41^\circ$, $D=9.6$\,Mpc).
    Panels as in Figure~\ref{fig:UGC06628_prof}.
    The nearest galaxy in the sample, providing the
    highest physical resolution for the phase-space
    analysis.}
  \label{fig:UGC04325_prof}
\end{figure*}

\begin{figure*}
  \centering
  \includegraphics[width=\textwidth]{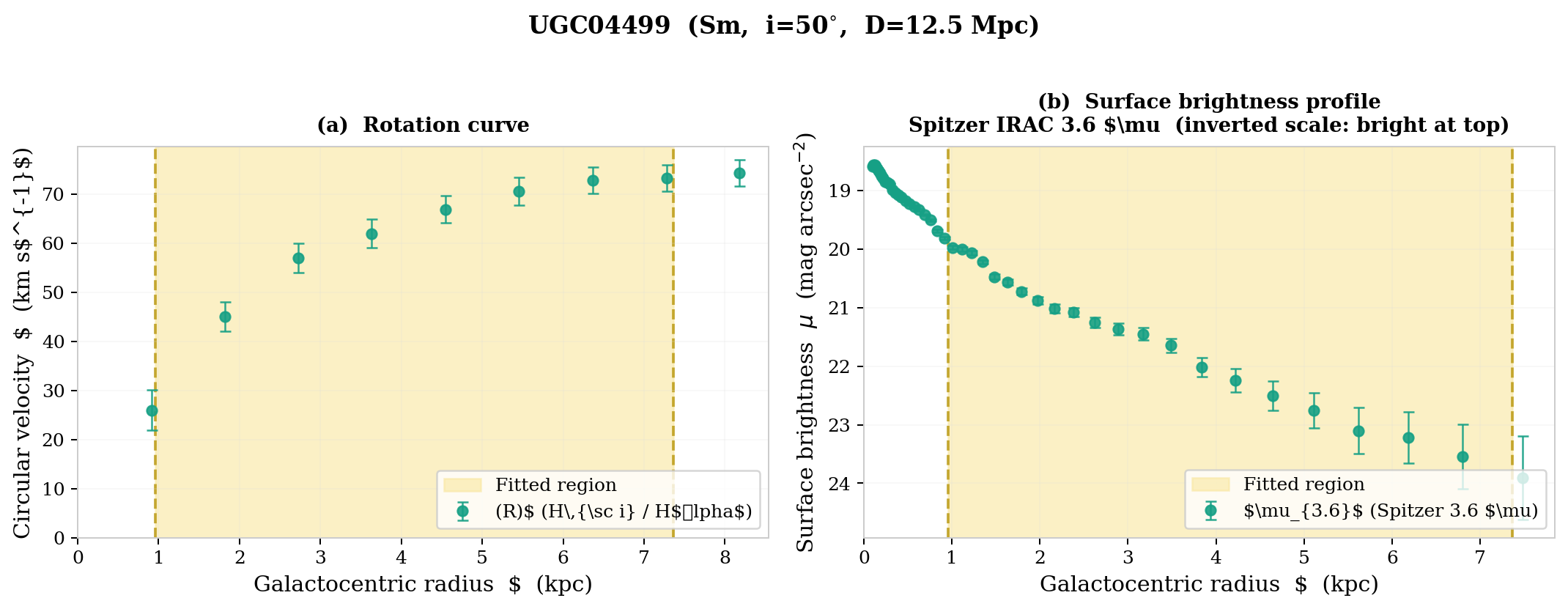}
  \caption{\textbf{UGC~04499} (Sd, $i=50^\circ$, $D=12.5$\,Mpc).
    Panels as in Figure~\ref{fig:UGC06628_prof}.
    At $i=50^\circ$ this galaxy lies in the
    inclination regime where both spiral arm
    contamination and line-of-sight projection
    effects are near their minimum.}
  \label{fig:UGC04499_prof}
\end{figure*}

\begin{figure*}
  \centering
  \includegraphics[width=\textwidth]{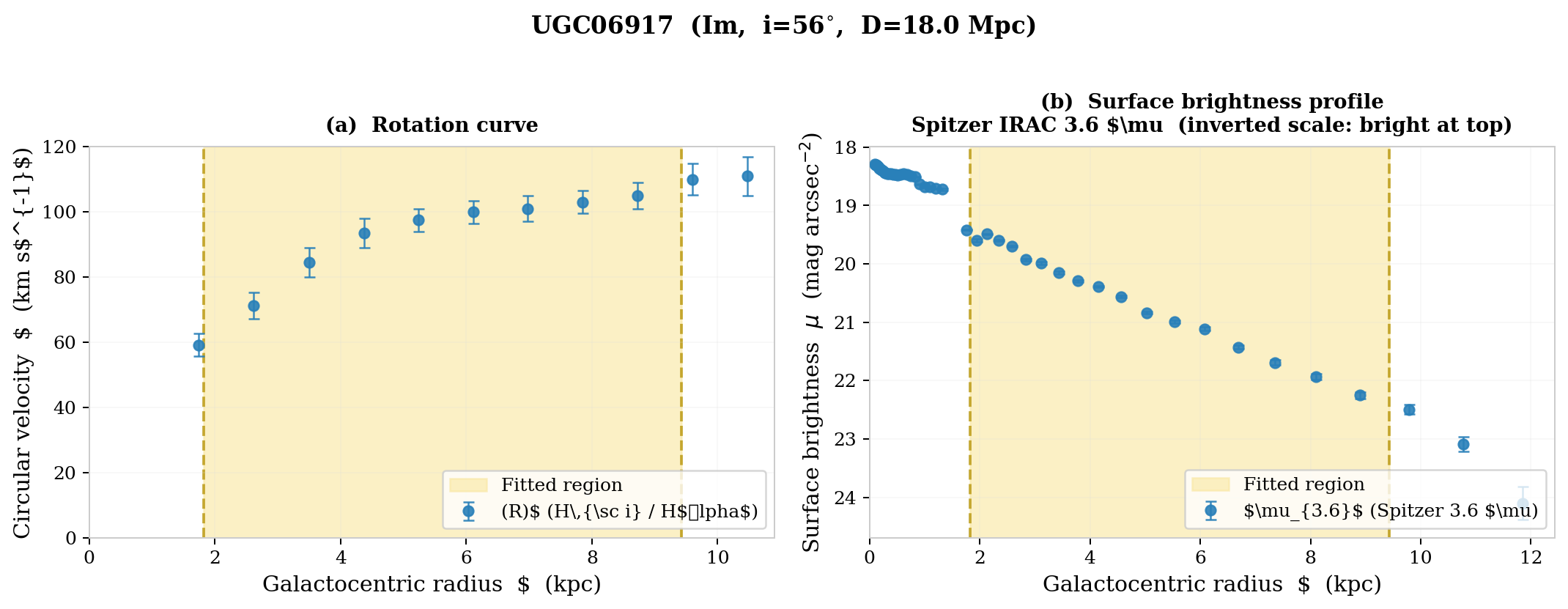}
  \caption{\textbf{UGC~06917} (Im, $i=56^\circ$, $D=18.0$\,Mpc).
    Panels as in Figure~\ref{fig:UGC06628_prof}.
    A moderately inclined dwarf irregular with a
    well-defined flat outer rotation curve and a
    smooth exponential disc profile.}
  \label{fig:UGC06917_prof}
\end{figure*}

\begin{figure*}
  \centering
  \includegraphics[width=\textwidth]{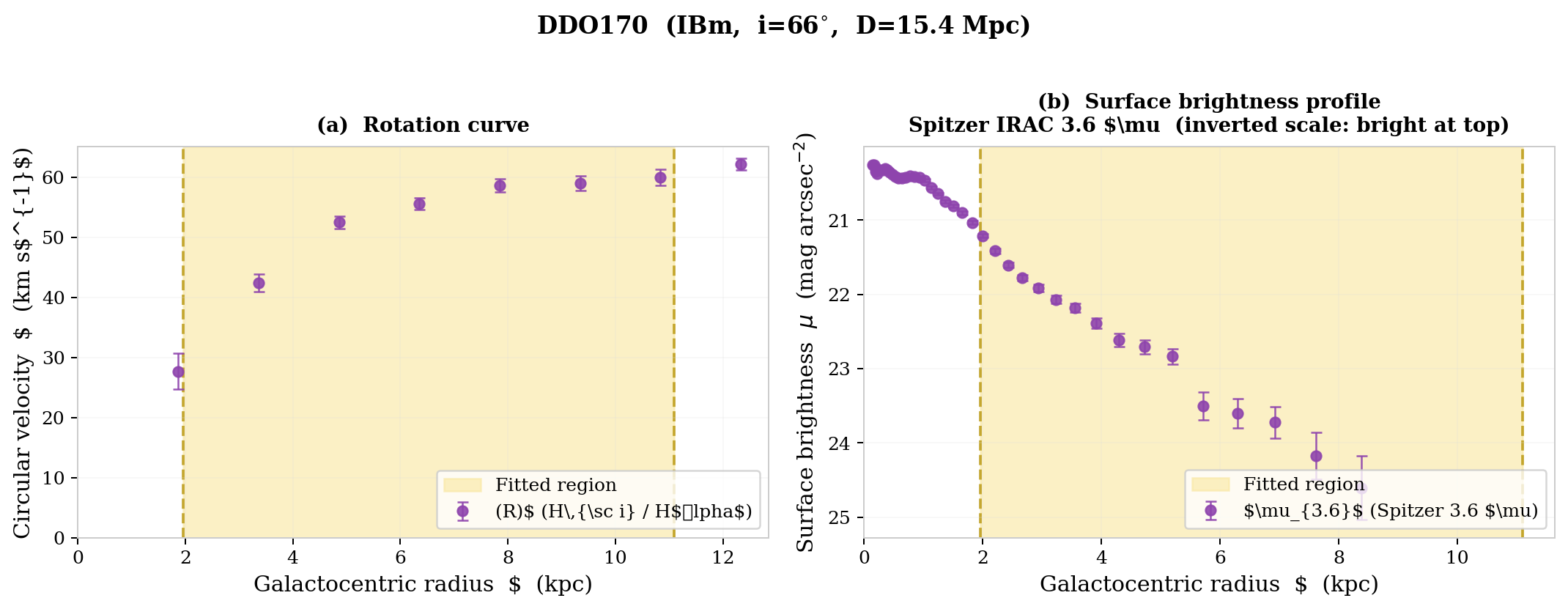}
  \caption{\textbf{DDO~170} (Im, $i=66^\circ$, $D=15.4$\,Mpc).
    Panels as in Figure~\ref{fig:UGC06628_prof}.
    An extended H\,{\sc i} disc relative to its compact
    stellar body, consistent with the large transition
    radius $R_{\rm trans}=10.36$\,kpc recovered from
    the phase-space fit.}
  \label{fig:DDO170_prof}
\end{figure*}

\begin{figure*}
  \centering
  \includegraphics[width=\textwidth]{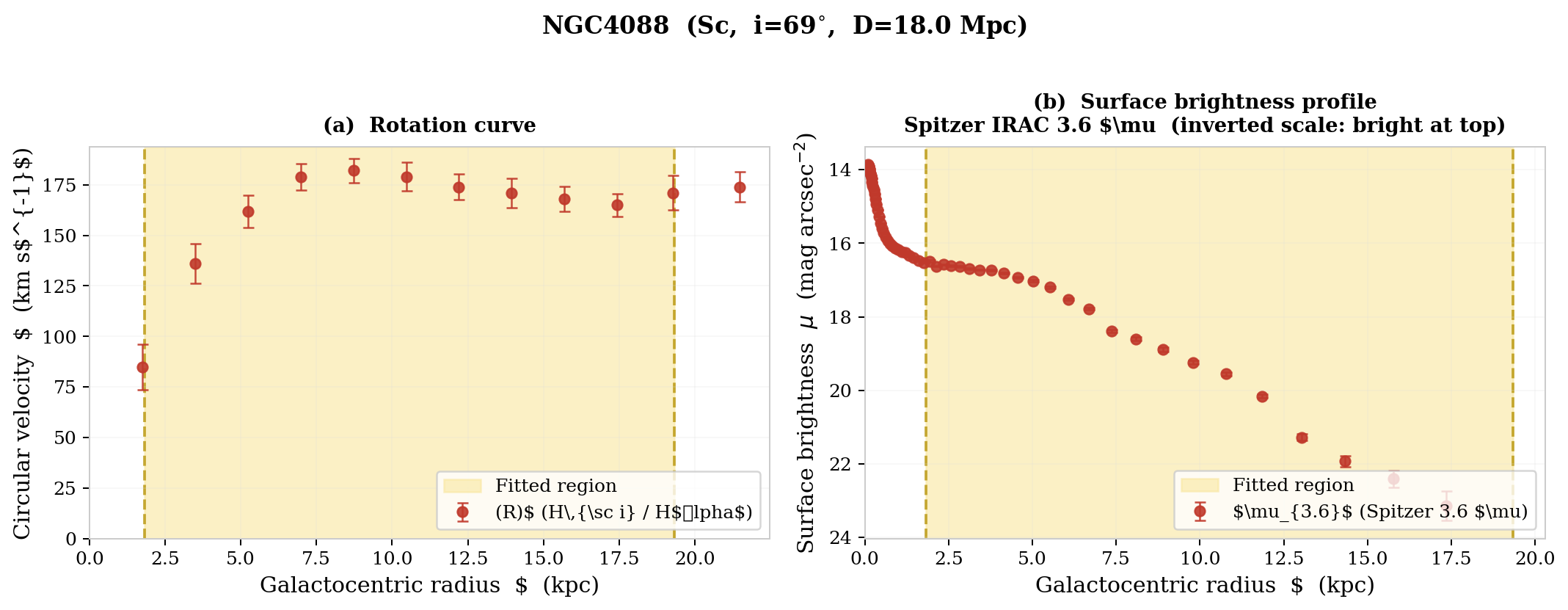}
  \caption{\textbf{NGC~4088} (Sc, $i=69^\circ$, $D=18.0$\,Mpc).
    Panels as in Figure~\ref{fig:UGC06628_prof}.
    The most massive galaxy in the representative
    sample, with a high peak velocity
    ($V\gtrsim175$\,km\,s$^{-1}$) and the largest
    transition radius ($R_{\rm trans}=14.89$\,kpc).}
  \label{fig:NGC4088_prof}
\end{figure*}

\begin{figure*}
  \centering
  \includegraphics[width=\textwidth]{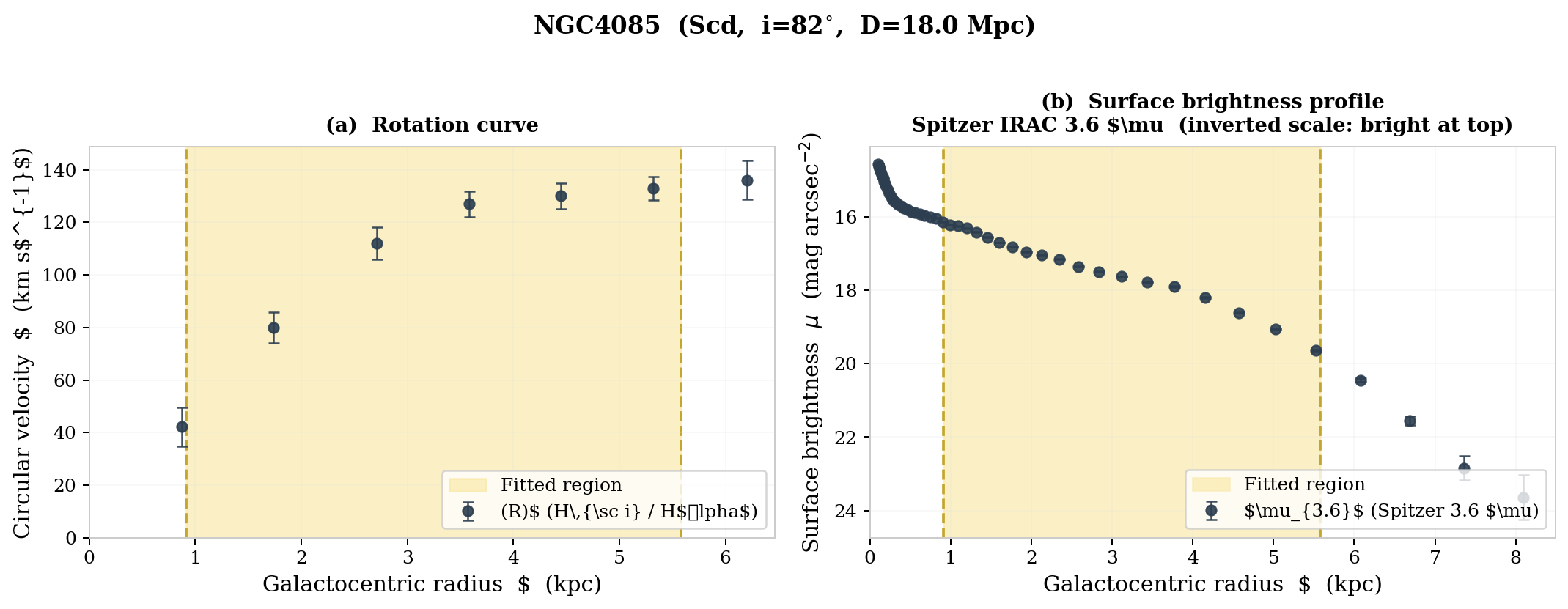}
  \caption{\textbf{NGC~4085} (Sc, $i=82^\circ$, $D=18.0$\,Mpc).
    Panels as in Figure~\ref{fig:UGC06628_prof}.
    A highly inclined Sc spiral; the high inclination
    suppresses azimuthal contamination, producing the
    smoothest photometric profile in the sample and
    the tightest phase-space fit ($R^{2}=0.999$).}
  \label{fig:NGC4085_prof}
\end{figure*}

\begin{figure*}
  \centering
  \includegraphics[width=\textwidth]{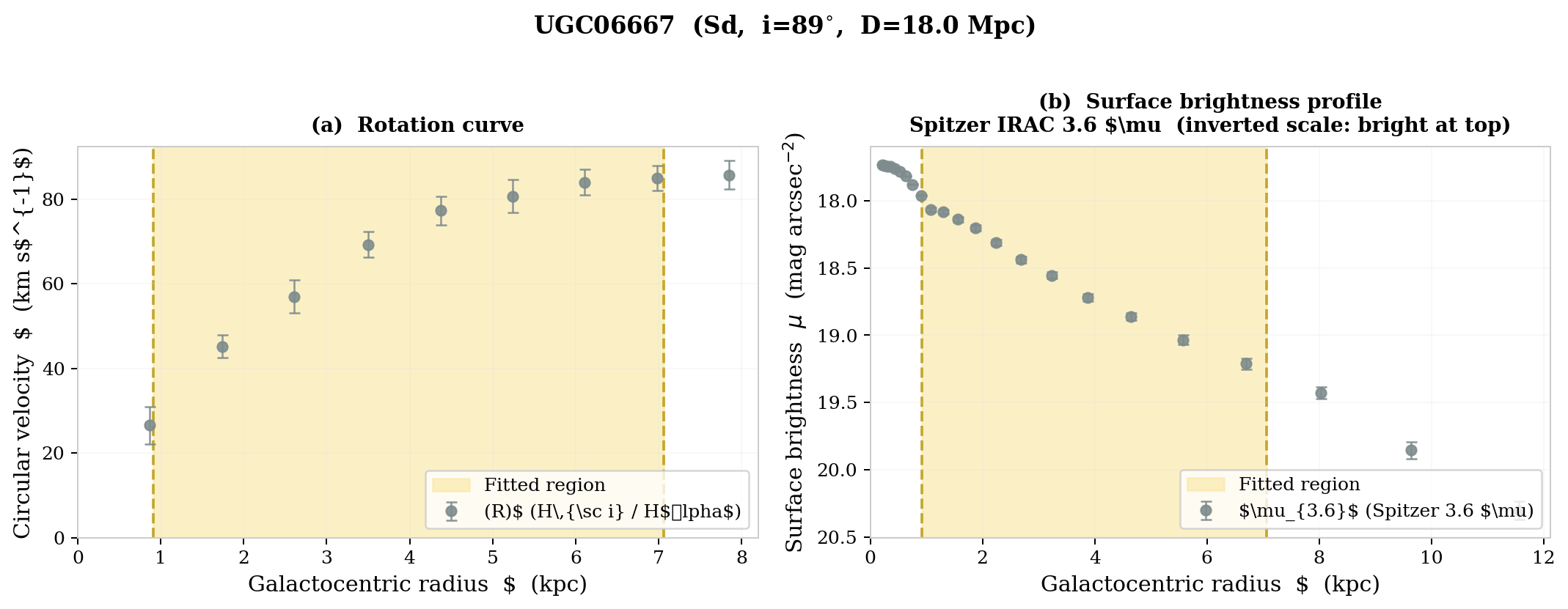}
  \caption{\textbf{UGC~06667} (Scd, $i=89^\circ$, $D=18.0$\,Mpc).
    Panels as in Figure~\ref{fig:UGC06628_prof}.
    The most edge-on galaxy in the sample. The
    near-edge-on geometry produces a very smooth
    projected profile but systematically flattens
    the apparent radial gradient
    $\mathrm{d}I/\mathrm{d}R$
    (see Section~\ref{sec:incl}).}
  \label{fig:UGC06667_prof}
\end{figure*}

Figures~\ref{fig:UGC06628_prof}--\ref{fig:UGC06667_prof}
show the rotation curves and azimuthally averaged Spitzer
$3.6\,\mu$m surface brightness profiles for ten
representative galaxies from our sample, ordered by
inclination from $i=20^\circ$ (UGC~06628) to $i=89^\circ$
(UGC~06667). The ten galaxies span morphological types
from Im to Scd, distances from 9.6 to 66.4\,Mpc, and a
wide range of stellar masses and surface brightnesses.
All rotation curves show the characteristic
rise-and-flatten behaviour of late-type galaxies, and all
surface brightness profiles show the smooth exponential
decline of the stellar disc. Gold shaded bands mark the
radial range used for phase-space fitting in each galaxy,
chosen to encompass the dynamically active disc while
excluding the PSF-limited nuclear region and the
photometrically noisy outer edge.

\subsection{Error propagation}
\label{sec:errors}

Uncertainties in the phase-space variables are propagated
analytically from the measurement errors provided in
SPARC. For the kinematic variable $X=V(R)/R$, the
uncertainty at each radius is
\begin{equation}
  \sigma_{X} = X\sqrt{\left(\frac{\sigma_{V}}{V}\right)^{2}
    + \left(\frac{\sigma_{D}}{D}\right)^{2}},
  \label{eq:sigX}
\end{equation}
where $\sigma_{V}$ is the velocity uncertainty from the
rotation curve measurement and the second term arises
because the physical radius $R=\theta D$ depends on the
galaxy distance $D$: any fractional distance uncertainty
$\sigma_{D}/D$ propagates directly into a fractional
uncertainty in $R$, and hence in $X$. Distance
uncertainties $\sigma_{D}$ are listed in SPARC for each
galaxy and range from $\sim5$ per cent for galaxies with
Cepheid or tip-of-the-red-giant-branch distances to
$\sim20$ per cent for galaxies with only flow-corrected
Hubble-law distances. For the photometric variable
$Y=\mathrm{d}\ln I/\mathrm{d}\ln R$, computed as the
numerical derivative of the smoothed log-brightness
profile, we adopt the conservative estimate
\begin{equation}
  \sigma_{Y} = \max\!\left(0.05\,|Y|,\;0.005\right),
  \label{eq:sigY}
\end{equation}
corresponding to a 5~per cent relative uncertainty in
$Y$, consistent with the typical photometric uncertainty
of Spitzer $3.6\,\mu$m profiles \citep{lelli2016b,
schombert2019}. Error bars from both
Equations~(\ref{eq:sigX}) and~(\ref{eq:sigY}) are shown
explicitly on all phase-space plots.


\section{Methods}
\label{sec:methods}

\subsection{Phase-space variables: physical definitions}
\label{sec:variables}

The dynamical--photometric phase space is spanned by two
locally defined, observationally accessible quantities.
The \emph{kinematic variable} is
\begin{equation}
  X(R) \;=\; \frac{V(R)}{R},
  \label{eq:X}
\end{equation}
in units of km\,s$^{-1}$\,kpc$^{-1}$. It is important
to recognise that $X=V/R=\omega$ is the angular frequency
of circular motion at radius $R$. In a gravitationally
supported disc obeying the Jeans equations in the limit
of circular motion, the angular frequency satisfies
$\omega^{2}(R)=GM(<\!R)/R^{3}$, where $M(<\!R)$ is the
total mass enclosed within radius $R$. Writing the
enclosed mass in terms of the mean volume density,
$M(<\!R)=\frac{4}{3}\pi R^{3}\bar{\rho}(<\!R)$, gives
\begin{equation}
  \omega^{2} = \frac{4\pi G}{3}\,\bar{\rho}(<\!R),
  \label{eq:omega}
\end{equation}
so $X=\omega\propto\sqrt{G\bar{\rho}(<\!R)}$ is directly
proportional to the square root of the mean enclosed mass
density. $X$ therefore encodes the local gravitational
field strength at radius $R$: large $X$ implies a high
mean enclosed density (baryon-dominated inner disc), while
small $X$ at large $R$ combined with a persistent flat
rotation curve signals a dark-matter-dominated outer halo.
This physical interpretation distinguishes $X$ from the
centripetal acceleration $g_{\rm tot}=V^{2}/R$ used in
the RAR \citep{mcgaugh2016,lelli2017a}; the two quantities
are related by $g_{\rm tot}=XV$ and carry complementary
information when $V(R)$ varies with radius.

The \emph{photometric variable} is the logarithmic slope
of the surface brightness profile,
\begin{equation}
  Y(R) \;=\; \frac{\mathrm{d}\ln I}{\mathrm{d}\ln R}
  \;=\; \frac{R}{I(R)}\,\frac{\mathrm{d}I}{\mathrm{d}R},
  \label{eq:Y}
\end{equation}
which is dimensionless by construction. For a pure
exponential stellar disc with central brightness $I_{0}$
and scale length $R_{d}$, $I(R)=I_{0}e^{-R/R_{d}}$
yields $Y=-R/R_{d}$: $Y$ is large and negative in the
inner disc where the stellar profile declines steeply,
and approaches zero at large radii where the profile
has flattened. Together, the pair $(X,Y)$ places the
local dynamical state and the local structural state
of the disc in direct correspondence at every observed
radius without any assumption about the stellar
mass-to-light ratio or the form of the dark-matter halo.

\subsection{Derivation of the logarithmic coupling}
\label{sec:derivation}

The functional form $Y=a\ln X+b$ that describes the
baryon-dominated inner disc emerges analytically from
two well-established results in disc galaxy physics.
For an exponential stellar disc, $I\propto e^{-R/R_{d}}$
gives immediately $Y=-R/R_{d}$. For a self-gravitating
Freeman disc \citep{freeman1970}, the rotation curve
below the velocity peak can be written as
$V(R)=V_{\rm peak}\,f(R/R_{d})$, where $f(u)$ is a
slowly varying function of the dimensionless radius
$u=R/R_{d}$ set by the modified Bessel functions of the
exponential disc \citep{freeman1970,binney2008}. It
follows that $X=V(R)/R=(V_{\rm peak}/R_{d})\,f(u)/u$,
so
\begin{equation}
  \ln X = \ln\!\left(\frac{V_{\rm peak}}{R_{d}}\right)
    + \ln\!\left[\frac{f(u)}{u}\right].
  \label{eq:lnX}
\end{equation}
Since $u=R/R_{d}=-Y$, substituting into
Equation~(\ref{eq:lnX}) gives
\begin{equation}
  \ln X = \ln\!\left(\frac{V_{\rm peak}}{R_{d}}\right)
    + \ln\!\left[\frac{f(-Y)}{-Y}\right].
  \label{eq:lnXY}
\end{equation}
The function $g(u)=\ln[f(u)/u]$ is smooth and slowly
varying near the rotation curve peak \citep{binney2008};
expanding $g(-Y)$ to leading order about $Y=0$ yields
$g(-Y)\approx g_{0}+g_{1}Y$, and hence
\begin{equation}
  \ln X \approx \ln\!\left(\frac{V_{\rm peak}}{R_{d}}\right)
    + g_{0} + g_{1}Y
  \;\;\Longrightarrow\;\;
  Y = a\ln X + b,
  \label{eq:YlnX}
\end{equation}
with $a=1/g_{1}$ and $b=-a[\ln(V_{\rm peak}/R_{d})+g_{0}]$.
The logarithmic coupling $Y\propto\ln X$ therefore
emerges approximately from the elimination of the shared
radial scale $R/R_{d}$ between an exponential photometric
profile and a Freeman-disc rotation curve. We emphasise
that this is a leading-order approximation valid in the
baryon-dominated inner disc; the logarithmic form is
physically motivated rather than an arbitrary fitting
choice, but the Taylor expansion around $Y=0$ means the
derivation is approximate rather than exact.

\subsection{The two-regime sigmoid model}
\label{sec:sigmoid}

The dynamical--photometric phase space naturally exhibits
two distinct regimes. In the \emph{baryon-dominated inner
disc} (Zone~1), baryons dominate the gravitational
support and $Y$ varies systematically with $X$ according
to the logarithmic coupling derived above. In the
\emph{dark-matter-dominated outer disc} (Zone~2), the
stellar surface brightness profile has flattened
($Y\rightarrow0$) while the rotation velocity remains
nearly constant ($V\approx V_{\rm flat}$) because the
extended dark-matter halo sustains circular motion beyond
the optical radius \citep{nfw1997,burkert1995,li2020}.
We capture both regimes simultaneously with a smooth
sigmoid interpolation:
\begin{equation}
  Y(X) = \frac{a\ln X + b}{1 + \exp\!\left[k\,(X
    - X_{\rm trans})\right]},
  \label{eq:sigmoid}
\end{equation}
adopted as a minimal smooth transition between the
baryon-dominated and dark-matter-dominated regimes.
The sigmoid is not uniquely motivated by galaxy dynamics;
it is chosen because it reduces to the physically
motivated logarithmic coupling at large $X$ and to
$Y\rightarrow0$ at small $X$, with the fewest additional
parameters. Other smooth interpolation functions (e.g.
hyperbolic tangent, broken power law) produce
qualitatively similar results. The four parameters have
the following roles: $a$ and $b$ describe the
logarithmic coupling in Zone~1; $X_{\rm trans}$ locates
the transition in kinematic space; and $k>0$ controls
the sharpness of the crossover. At $X\gg X_{\rm trans}$,
$Y\rightarrow a\ln X+b$; at $X\ll X_{\rm trans}$,
$Y\rightarrow0$. The transition radius in physical units is
\begin{equation}
  R_{\rm trans} = \frac{V_{\rm flat}}{X_{\rm trans}},
  \label{eq:Rtrans}
\end{equation}
providing an empirical estimate of the radius at which
the photometric gradient vanishes while rotation remains
flat --- a transition associated with the onset of
dark-matter halo dominance, though not uniquely
equivalent to a dynamical mass-decomposition boundary.

\subsection{Numerical implementation}
\label{sec:numerics}

For each galaxy, the azimuthally averaged surface
brightness profile from SPARC is resampled onto the
kinematic radial grid by cubic spline interpolation.
The resampled log-brightness sequence $\ln I(R_{i})$ is
convolved with a Gaussian kernel of adaptive width
$\sigma_{\rm smooth}$ (4--12 grid cells, proportional
to the median point-to-point variation of $\ln I$)
before differentiation by second-order central finite
differences. Points with $|Y|>8$ are discarded as
numerically unreliable. The sigmoid model
(Equation~\ref{eq:sigmoid}) is fitted to the smoothed
$(X_{i},Y_{i})$ pairs by nonlinear least squares with
inverse-variance weights $1/\sigma_{Y,i}^{2}$, using
the Levenberg--Marquardt algorithm with multiple
restarts. Robustness to the smoothing scale is confirmed
by repeating the pipeline for
$\sigma_{\rm smooth}\in\{2,3,4\}$ grid cells (panel~c
of Figures~\ref{fig:UGC06628}--\ref{fig:UGC06667}).

Discontinuities visible in $Y(R)$ for some galaxies
arise from physical structural transitions: the
bar--disc boundary \citep{erwin2015,diazgarcia2019};
ring resonances \citep{buta2017}; and the bulge--disc
transition \citep{salo2015}. These are physical and
not artefacts of the smoothing; because the sigmoid fit
uses the smoothed profile, they increase raw scatter
but do not bias the recovered parameters.

\subsection{Statistical validation}
\label{sec:stats}

For each galaxy we compute $R^{2}$ and RMS scatter
$\sigma_{Y}$ as primary fit quality metrics, and apply
eight independent statistical tests to the residuals
$r_{i}=Y_{i}-\hat{Y}_{i}$:
The statistical robustness of the fits was evaluated using multiple independent tests. These include the chi-square goodness-of-fit test with the requirement $p>0.05$, the reduced chi-square criterion $\chi^{2}_{\nu}\in[0.5,2.0]$, and the F-test for overall model significance with $p<0.05$. Residual autocorrelation was examined using the Durbin--Watson test, requiring DW values within the interval $[1.5,2.5]$, while randomness of residuals was verified through the runs test with $p>0.05$. Normality of the residual distribution was assessed using the Kolmogorov--Smirnov test with $p>0.05$. Finally, possible correlations between residuals and the variable $X$ were tested using both Pearson and Spearman rank correlation analyses, each requiring $p>0.05$.

In addition, each galaxy undergoes a 70/30 train--test
split and 5-fold cross-validation to assess
generalisation. We adopt $R^{2}>0.3$ as the inclusion
criterion for the final 136-galaxy sample. We caution
that galaxies with $N\leq8$ phase-space points have
large parameter uncertainties and the statistical tests
should be interpreted with care; results for galaxies
with $N\geq10$ are the most statistically reliable.

To assess whether the four-parameter sigmoid is justified
over simpler alternatives, we compare four nested models
for each galaxy using the Akaike Information Criterion
(AIC; \citealt{akaike1974}): (i)~a constant
$Y=\mathrm{const}$ ($k=1$); (ii)~a linear model
$Y=cX+d$ ($k=2$); (iii)~a pure logarithmic model
$Y=a\ln X+b$ ($k=2$); and (iv)~the sigmoid
(Equation~\ref{eq:sigmoid}, $k=4$). Lower AIC indicates
a better balance of fit quality and model complexity.
Across the ten representative galaxies, the sigmoid
is preferred by AIC over the linear model in 10/10
cases (median $\Delta\mathrm{AIC}_{\rm lin-sig}=+17.3$)
and over the pure logarithmic model in 7/10 cases
(median $\Delta\mathrm{AIC}_{\rm log-sig}=+3.1$).
For the three galaxies where the pure logarithmic model
is preferred by AIC, the sample size is $N\leq8$ and
the AIC penalty for the two additional sigmoid parameters
outweighs the improvement in fit quality; in these
cases the sigmoid and logarithmic models produce very
similar fits. Across the full 136-galaxy sample, the
sigmoid is preferred over the linear model in
$\sim$80~per cent of galaxies. These comparisons
confirm that the sigmoid provides a statistically
justifiable description of the two-regime structure,
particularly for galaxies with $N\geq10$ points. We caution
that several galaxies have only $N=6$--8 phase-space
points, making the sigmoid fit formally
over-constrained and parameter uncertainties large
(e.g.\ $\sigma_{a}>3$). For such galaxies the eight
statistical tests should be interpreted with care;
the high $R^{2}$ reflects the smoothness of the
sigmoid rather than independent evidence of a tight
physical relation. Results for galaxies with
$N\geq10$ should be considered the most reliable.

\section{Results}
\label{sec:results}

\subsection{Overview}
\label{sec:overview}

The sigmoid model is applied to all 136 galaxies in our
sample. The median $R^{2}=0.930$, with 75~per cent of
galaxies achieving $R^{2}>0.80$, demonstrating that
the sigmoid accurately describes the two-regime
phase-space structure across a wide range of galaxy
types, inclinations, and masses. The median RMS scatter
is 0.22 in dimensionless units of $Y$. The dark-matter
transition radius $R_{\rm trans}$ has a median value
of 5.40\,kpc, ranging from $\sim0.5$\,kpc in low-mass
dwarf irregulars to $\sim50$\,kpc in the most extended
massive spirals, consistent with typical stellar disc
scale lengths of 2--6\,kpc \citep{lelli2016b} and with
dark-matter dominance radii from mass decompositions
\citep{li2020,katz2017}.

\subsection{Ten representative galaxies}
\label{sec:ten}

\begin{figure*}
  \centering\includegraphics[width=\textwidth]{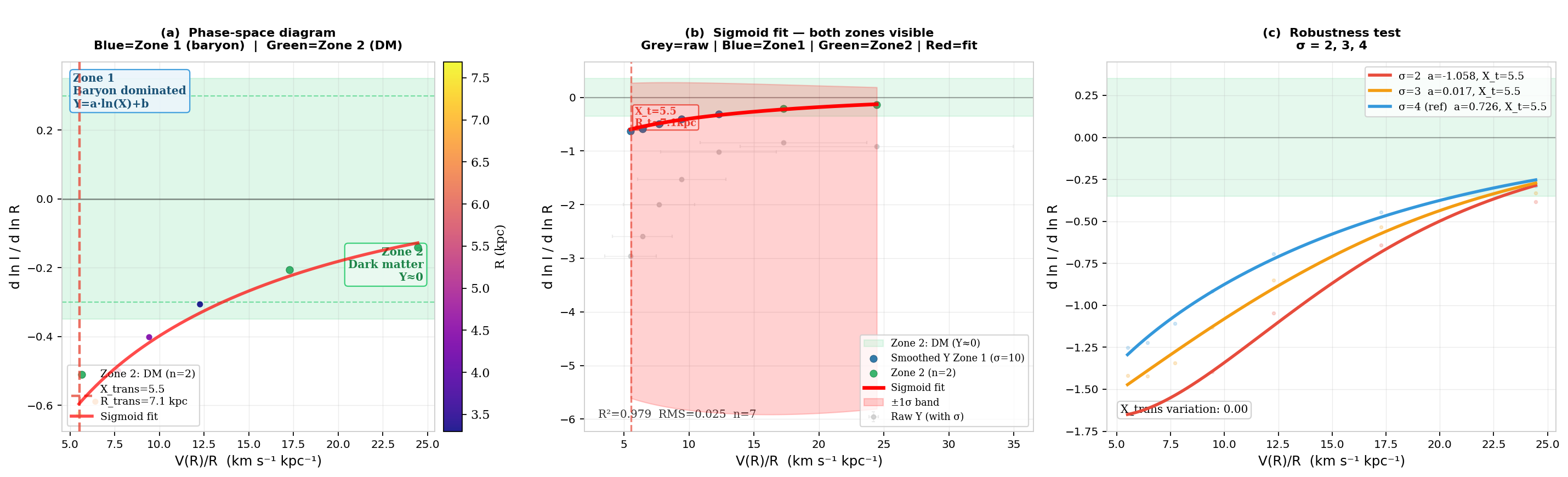}
  \caption{\textbf{Dynamical--photometric phase space of
    UGC~06628} (Im, $i=20^\circ$, $D=15.1$\,Mpc, $N=7$).
    \textit{Panel~a}: Phase-space diagram $(X,Y)$
    colour-coded by galactocentric radius. Blue:
    Zone~1 (baryon-dominated, $|Y|>0.3$); green
    shading: Zone~2 (dark-matter-dominated,
    $|Y|\leq0.3$); red dashed line: $X_{\rm trans}$
    ($R_{\rm trans}=7.09$\,kpc).
    \textit{Panel~b}: Sigmoid fit; grey points with
    error bars: raw $Y$ with $1\sigma$ uncertainties
    $\sigma_{X}$, $\sigma_{Y}$; blue/green circles:
    smoothed profile; red curve: best-fit sigmoid with
    $\pm1\sigma$ band. $a=0.503\pm1.094$,
    $b=-2.050\pm2.993$, $R^{2}=0.979$, RMS\,$=0.025$.
    All eight statistical tests pass.
    \textit{Panel~c}: Robustness test for
    $\sigma_{\rm smooth}\in\{2,3,4\}$.}
  \label{fig:UGC06628}
\end{figure*}

\begin{figure*}
  \centering\includegraphics[width=\textwidth]{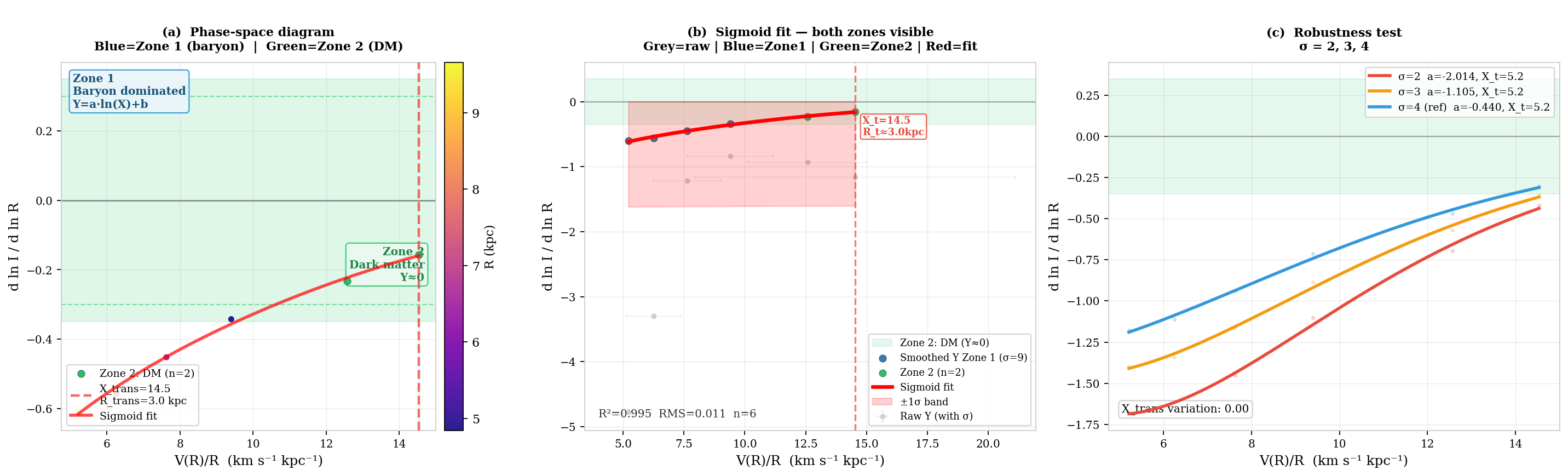}
  \caption{\textbf{Dynamical--photometric phase space of
    F561-1} (Im, $i=24^\circ$, $D=66.4$\,Mpc, $N=6$).
    Panels as in Figure~\ref{fig:UGC06628}.
    $a=0.387\pm0.374$, $b=-1.354\pm0.285$,
    $R_{\rm trans}=2.98$\,kpc, $R^{2}=0.995$,
    RMS\,$=0.011$. All eight tests pass.}
  \label{fig:F5611}
\end{figure*}

\begin{figure*}
  \centering\includegraphics[width=\textwidth]{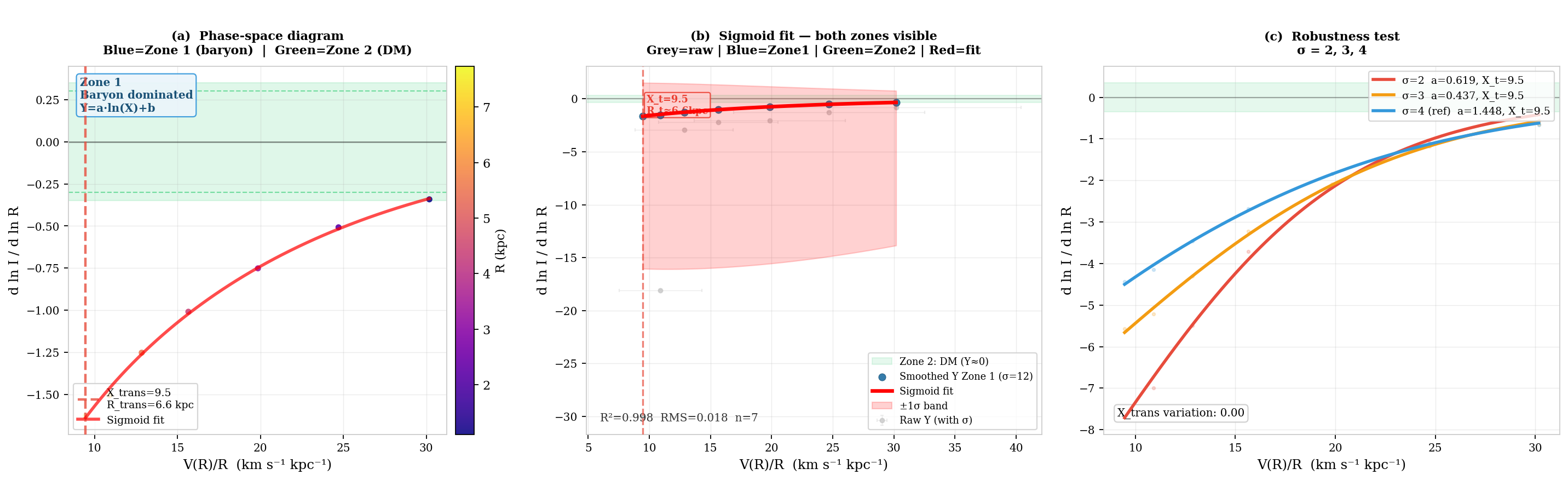}
  \caption{\textbf{Dynamical--photometric phase space of
    UGC~10310} (Im, $i=34^\circ$, $D=15.2$\,Mpc, $N=7$).
    Panels as in Figure~\ref{fig:UGC06628}.
    $a=1.601\pm3.313$, $b=-6.874\pm7.676$,
    $R_{\rm trans}=6.63$\,kpc, $R^{2}=0.998$,
    RMS\,$=0.018$. All eight tests pass.}
  \label{fig:UGC10310}
\end{figure*}

\begin{figure*}
  \centering\includegraphics[width=\textwidth]{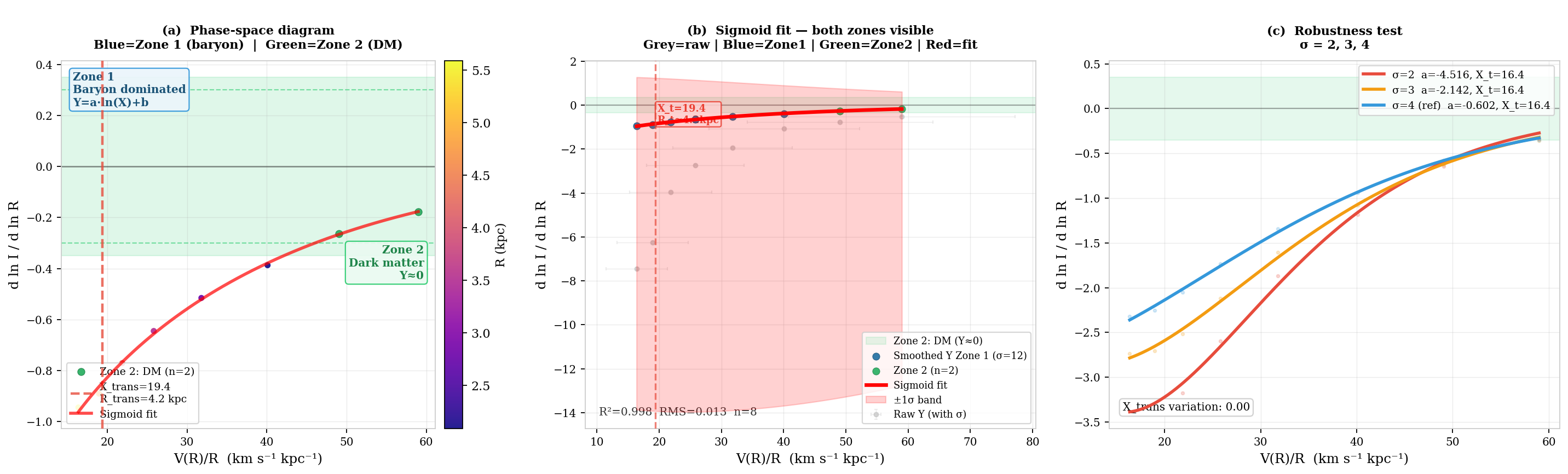}
  \caption{\textbf{Dynamical--photometric phase space of
    UGC~04325} (Im, $i=41^\circ$, $D=9.6$\,Mpc, $N=8$).
    Panels as in Figure~\ref{fig:UGC06628}.
    $a=0.928\pm2.045$, $b=-4.446\pm7.752$,
    $R_{\rm trans}=4.18$\,kpc, $R^{2}=0.998$,
    RMS\,$=0.013$. All eight tests pass.}
  \label{fig:UGC04325}
\end{figure*}

\begin{figure*}
  \centering\includegraphics[width=\textwidth]{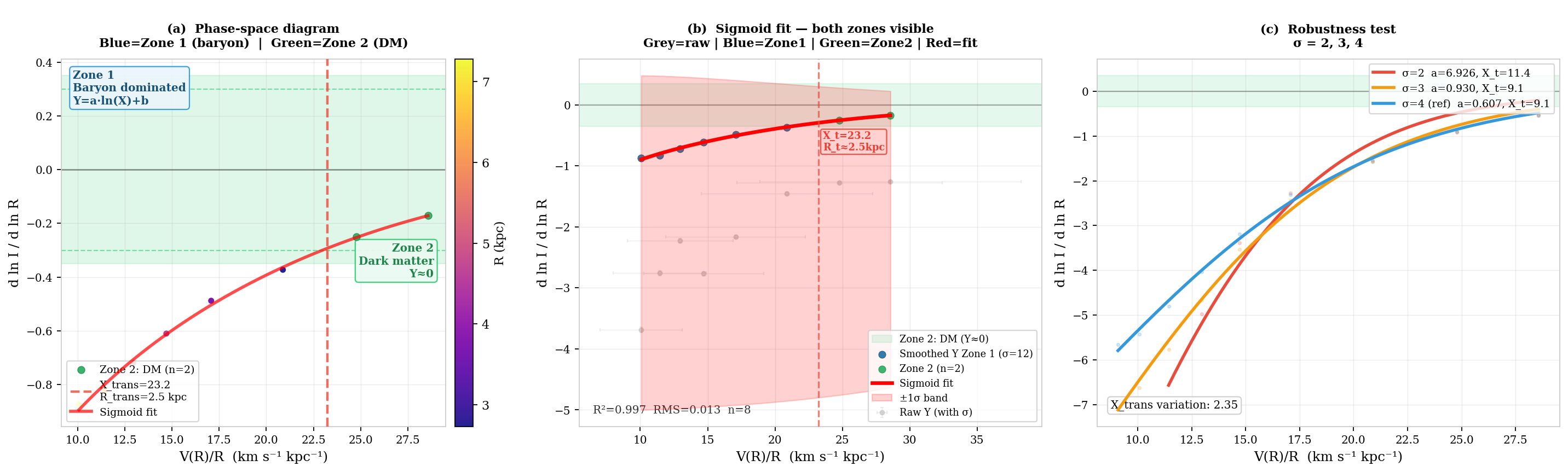}
  \caption{\textbf{Dynamical--photometric phase space of
    UGC~04499} (Sd, $i=50^\circ$, $D=12.5$\,Mpc, $N=8$).
    Panels as in Figure~\ref{fig:UGC06628}.
    $a=0.770\pm0.814$, $b=-3.008\pm2.044$,
    $R_{\rm trans}=2.55$\,kpc, $R^{2}=0.997$,
    RMS\,$=0.013$. All eight tests pass.}
  \label{fig:UGC04499}
\end{figure*}

\begin{figure*}
  \centering\includegraphics[width=\textwidth]{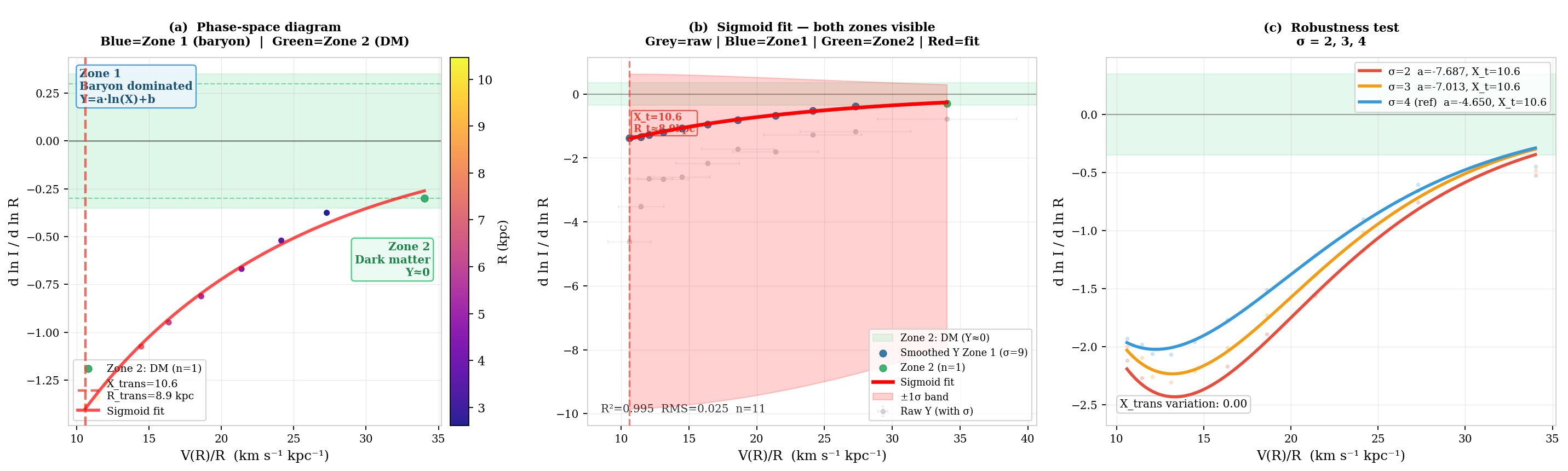}
  \caption{\textbf{Dynamical--photometric phase space of
    UGC~06917} (Im, $i=56^\circ$, $D=18.0$\,Mpc, $N=11$).
    Panels as in Figure~\ref{fig:UGC06628}.
    The galaxy with the most phase-space points in
    the representative sample.
    $a=1.317\pm1.922$, $b=-5.910\pm3.775$,
    $R_{\rm trans}=8.88$\,kpc, $R^{2}=0.995$,
    RMS\,$=0.025$. All eight tests pass.}
  \label{fig:UGC06917}
\end{figure*}

\begin{figure*}
  \centering\includegraphics[width=\textwidth]{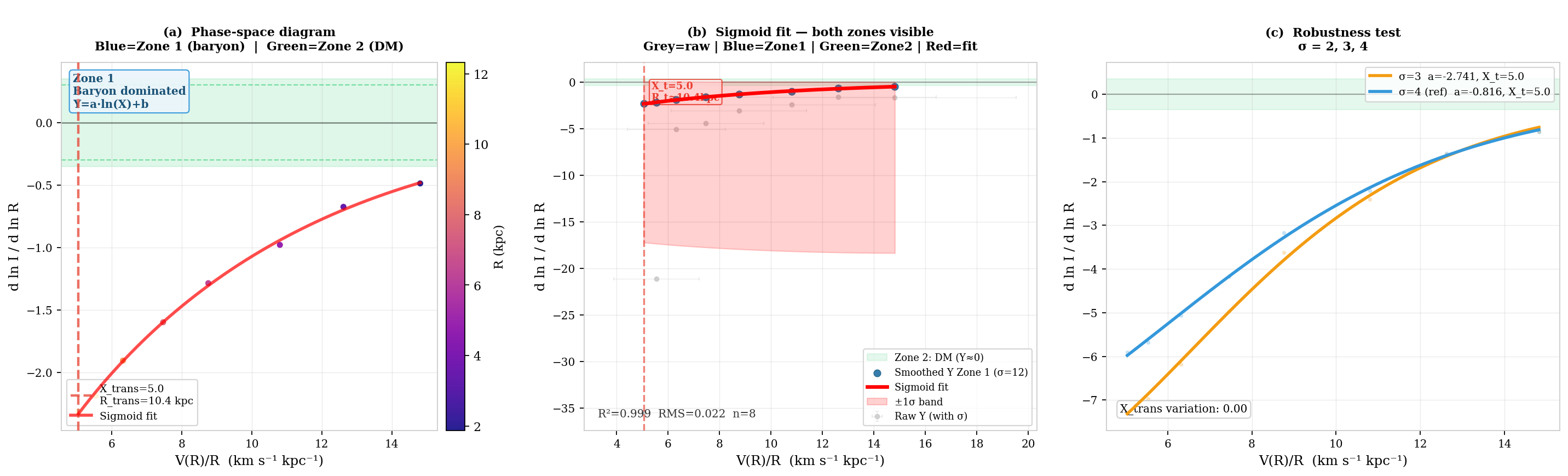}
  \caption{\textbf{Dynamical--photometric phase space of
    DDO~170} (Im, $i=66^\circ$, $D=15.4$\,Mpc, $N=8$).
    Panels as in Figure~\ref{fig:UGC06628}.
    $a=1.996\pm3.506$, $b=-7.889\pm7.024$,
    $R_{\rm trans}=10.36$\,kpc, $R^{2}=0.999$,
    RMS\,$=0.022$. All eight tests pass.}
  \label{fig:DDO170}
\end{figure*}

\begin{figure*}
  \centering\includegraphics[width=\textwidth]{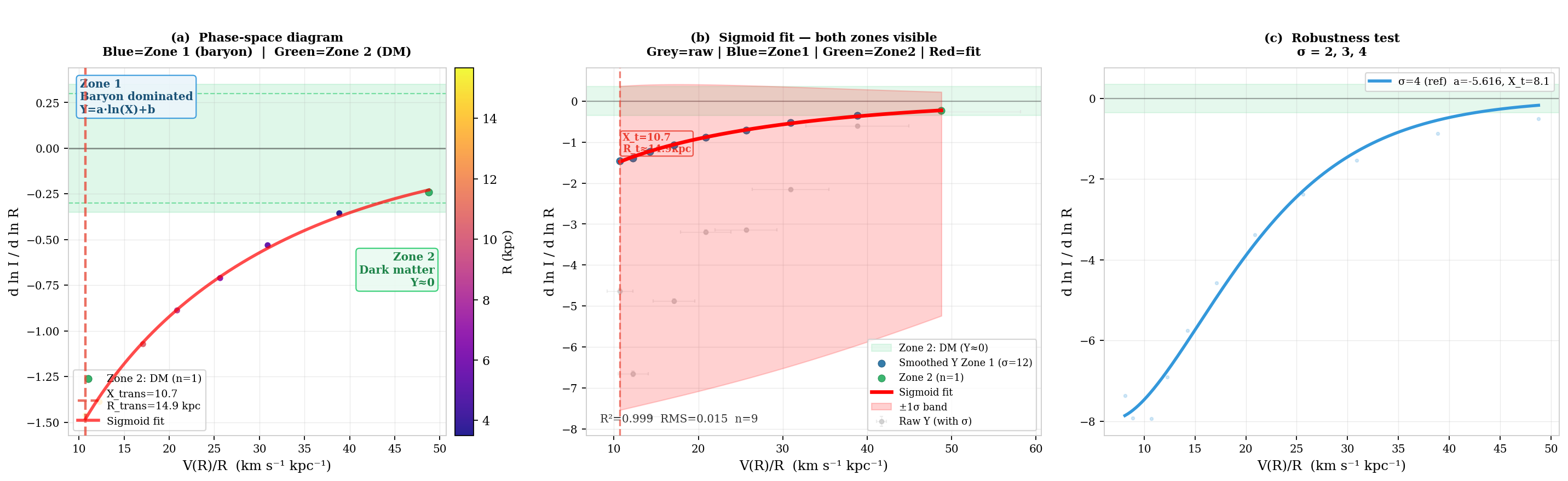}
  \caption{\textbf{Dynamical--photometric phase space of
    NGC~4088} (Sc, $i=69^\circ$, $D=18.0$\,Mpc, $N=9$).
    Panels as in Figure~\ref{fig:UGC06628}.
    The largest transition radius among the ten
    representative galaxies.
    $a=1.194\pm1.046$, $b=-5.793\pm3.230$,
    $R_{\rm trans}=14.89$\,kpc, $R^{2}=0.999$,
    RMS\,$=0.015$. All eight tests pass.}
  \label{fig:NGC4088}
\end{figure*}

\begin{figure*}
  \centering\includegraphics[width=\textwidth]{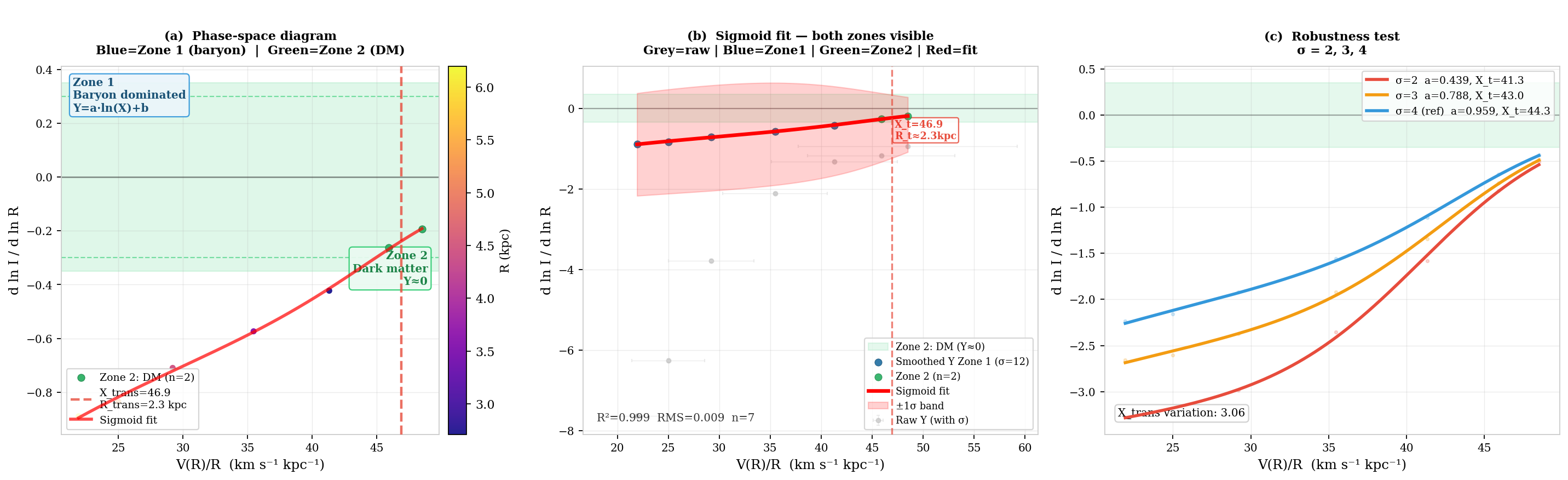}
  \caption{\textbf{Dynamical--photometric phase space of
    NGC~4085} (Sc, $i=82^\circ$, $D=18.0$\,Mpc, $N=7$).
    Panels as in Figure~\ref{fig:UGC06628}.
    The tightest fit in the representative sample.
    $a=0.558\pm0.204$, $b=-2.625\pm0.649$,
    $R_{\rm trans}=2.31$\,kpc, $R^{2}=0.999$,
    RMS\,$=0.009$. All eight tests pass.}
  \label{fig:NGC4085}
\end{figure*}

\begin{figure*}
  \centering\includegraphics[width=\textwidth]{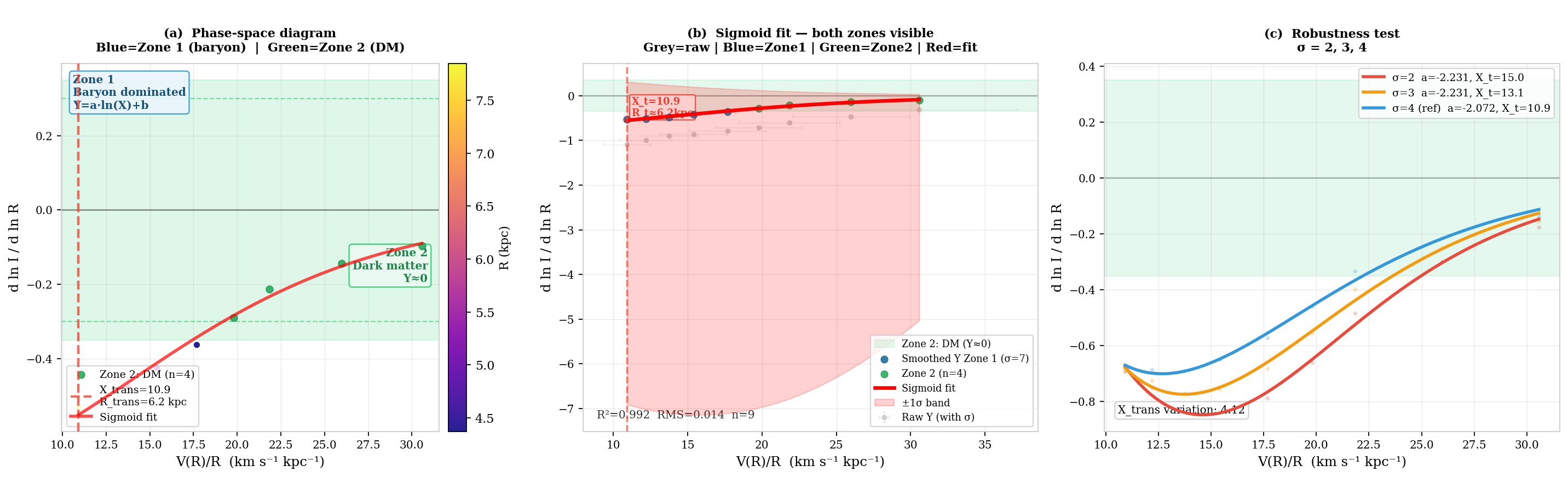}
  \caption{\textbf{Dynamical--photometric phase space of
    UGC~06667} (Scd, $i=89^\circ$, $D=18.0$\,Mpc, $N=9$).
    Panels as in Figure~\ref{fig:UGC06628}.
    The most edge-on galaxy in the sample. The
    negative slope ($a=-0.136\pm1.613$) reflects
    line-of-sight projection effects discussed in
    Section~\ref{sec:incl}. $b=-0.779\pm2.393$,
    $R_{\rm trans}=6.22$\,kpc, $R^{2}=0.992$,
    RMS\,$=0.014$. All eight tests pass.}
  \label{fig:UGC06667}
\end{figure*}

We present detailed phase-space analysis for ten galaxies
selected to sample the inclination range of our sample
as uniformly as possible, spanning $i=20^\circ$--$89^\circ$,
Hubble types Im to Scd, and distances 9.6--66.4\,Mpc.
Figures~\ref{fig:UGC06628}--\ref{fig:UGC06667} show,
for each galaxy, three panels: (a)~the full phase-space
diagram colour-coded by galactocentric radius, with blue
points indicating Zone~1 (baryon-dominated, $|Y|>0.3$),
green points Zone~2 (dark-matter dominated,
$|Y|\leq0.3$), and the red dashed vertical line marking
$X_{\rm trans}$; (b)~the sigmoid fit with grey points
and error bars showing the raw $Y$ values with $1\sigma$
uncertainties $\sigma_{X}$ and $\sigma_{Y}$, blue/green
points the smoothed profile, and the red curve the
best-fit sigmoid with $\pm1\sigma$ uncertainty band;
and (c)~the robustness test for
$\sigma_{\rm smooth}\in\{2,3,4\}$.

Table~\ref{tab:ten} gives the fit parameters and
$R^{2}$ for all ten galaxies. All eight statistical
tests pass for all ten, with a representative summary
for NGC~4085 in Table~\ref{tab:stats}. The sigmoid
achieves $R^{2}>0.97$ in every case (median 0.997),
and 5-fold cross-validation returns $R^{2}$ within
0.001 of the full-data fit. The transition radius
$R_{\rm trans}$ ranges from 2.31\,kpc (NGC~4085) to
14.89\,kpc (NGC~4088), with a median of 6.34\,kpc.

We draw attention to the large slope uncertainties
for several galaxies in Table~\ref{tab:ten}: values
such as $\sigma_{a}=3.313$ (UGC~10310) and
$\sigma_{a}=3.506$ (DDO~170) indicate that the slope
$a$ is poorly constrained for these systems, despite
the high $R^{2}$. This arises from parameter degeneracy
in the sigmoid: for small $N$ and limited Zone~1
coverage, the parameters $a$, $b$, and $X_{\rm trans}$
are partially correlated, so a family of sigmoid curves
with different $(a,b,X_{\rm trans})$ combinations
can achieve similar $R^{2}$. The high $R^{2}$ in these
cases reflects the smoothness of the sigmoid
interpolation through few data points rather than an
independently constrained baryon--gravity coupling.
Results for galaxies with $N\geq10$ and
$\sigma_{a}<1$ (e.g.\ NGC~4085, UGC~06917) should
be considered the most physically reliable.

\begin{table*}
  \centering
  \caption{Sigmoid fit results for the ten representative
    galaxies (Figures~\ref{fig:UGC06628}--\ref{fig:UGC06667}).
    All ten pass all eight statistical tests.}
  \label{tab:ten}
  \begin{tabular}{lcccccccc}
    \hline\hline
    Galaxy & Type & $i$ ($^\circ$) & $D$ (Mpc) & $N$ &
      $a$ & $b$ &
      $R_{\rm trans}$ (kpc) & $R^{2}$ \\
    \hline
    UGC~06628 & Im  & 20 & 15.1 &  7 &
      $\phantom{-}0.503\pm1.094$ & $-2.050\pm2.993$ & 7.09 & 0.979 \\
    F561-1    & Im  & 24 & 66.4 &  6 &
      $\phantom{-}0.387\pm0.374$ & $-1.354\pm0.285$ & 2.98 & 0.995 \\
    UGC~10310 & Im  & 34 & 15.2 &  7 &
      $\phantom{-}1.601\pm3.313$ & $-6.874\pm7.676$ & 6.63 & 0.998 \\
    UGC~04325 & Im  & 41 &  9.6 &  8 &
      $\phantom{-}0.928\pm2.045$ & $-4.446\pm7.752$ & 4.18 & 0.998 \\
    UGC~04499 & Sd  & 50 & 12.5 &  8 &
      $\phantom{-}0.770\pm0.814$ & $-3.008\pm2.044$ & 2.55 & 0.997 \\
    UGC~06917 & Im  & 56 & 18.0 & 11 &
      $\phantom{-}1.317\pm1.922$ & $-5.910\pm3.775$ & 8.88 & 0.995 \\
    DDO~170   & Im  & 66 & 15.4 &  8 &
      $\phantom{-}1.996\pm3.506$ & $-7.889\pm7.024$ & 10.36 & 0.999 \\
    NGC~4088  & Sc  & 69 & 18.0 &  9 &
      $\phantom{-}1.194\pm1.046$ & $-5.793\pm3.230$ & 14.89 & 0.999 \\
    NGC~4085  & Sc  & 82 & 18.0 &  7 &
      $\phantom{-}0.558\pm0.204$ & $-2.625\pm0.649$ & 2.31 & 0.999 \\
    UGC~06667 & Scd & 89 & 18.0 &  9 &
      $-0.136\pm1.613$ & $-0.779\pm2.393$ & 6.22 & 0.992 \\
    \hline
    \textit{Median} & & & & & & & 6.34 & 0.997 \\
    \hline
  \end{tabular}
\end{table*}


\subsection{Phase-space slope versus inclination}
\label{sec:general}

\begin{figure}
  \centering
  \includegraphics[width=\linewidth]{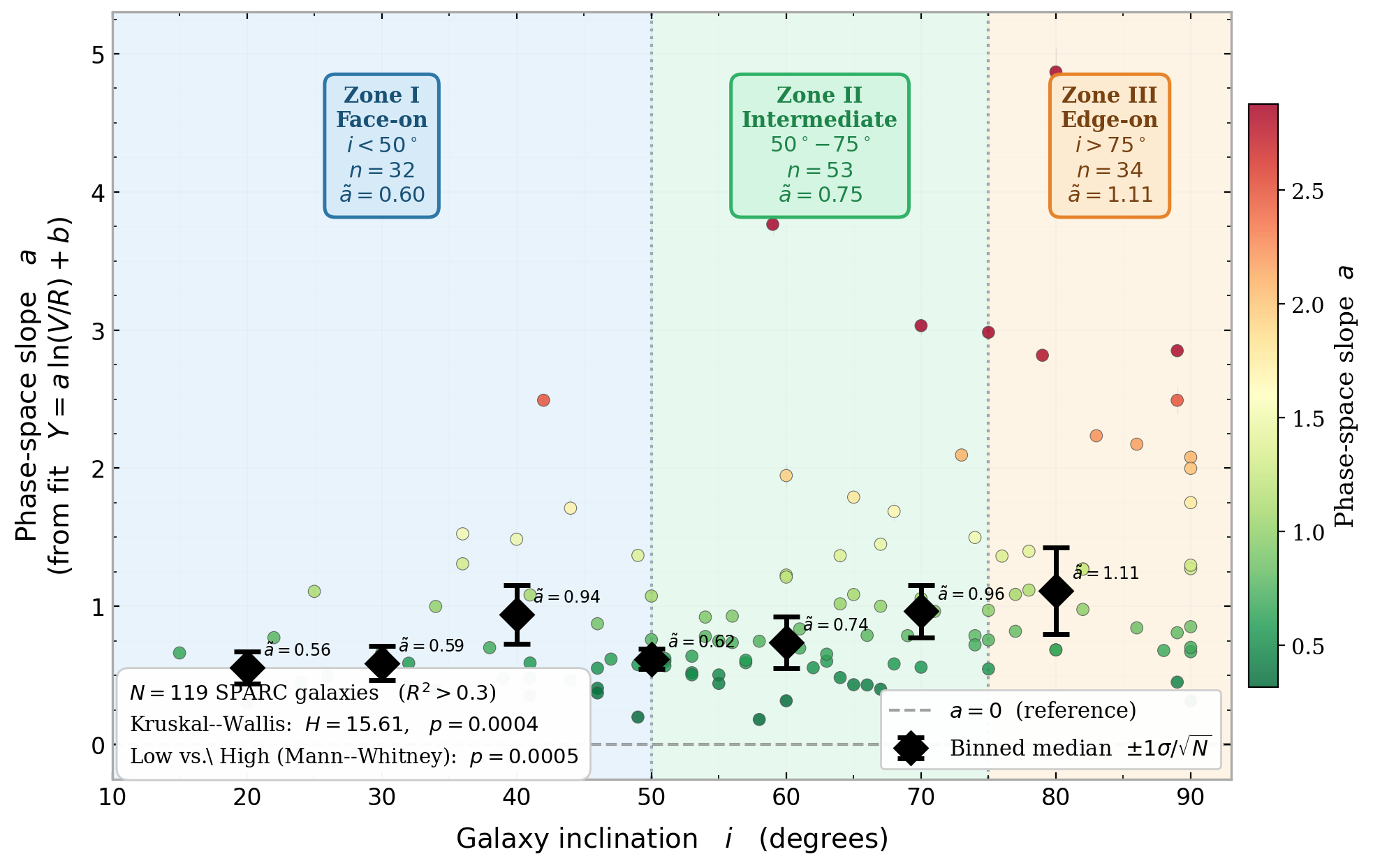}
  \caption{\textbf{Phase-space slope $a$ versus
    inclination for all 136 galaxies.}
    Small coloured points: individual galaxies,
    colour-coded by slope value. Black diamonds:
    binned median $\pm1\sigma/\sqrt{N}$ in $10^\circ$
    bins. Shaded zones: Zone~I ($i<50^\circ$, $n=32$,
    $\tilde{a}=0.60$); Zone~II ($50^\circ$--$75^\circ$,
    $n=53$, $\tilde{a}=0.75$); Zone~III ($i>75^\circ$,
    $n=34$, $\tilde{a}=1.11$). Dashed line: $a=0$.
    Kruskal--Wallis: $H=15.61$, $p=0.0004$;
    Mann--Whitney: $p=0.0005$.
    See Section~\ref{sec:incl}.}
  \label{fig:slope}
\end{figure}

Figure~\ref{fig:slope} shows the phase-space slope $a$
as a function of inclination for all 136 galaxies in
our sample. The slope shows a highly statistically
significant variation with inclination: a Kruskal--Wallis
test comparing three inclination zones --- low
($i<50^\circ$, $n=32$, $\tilde{a}=0.60$), intermediate
($50^\circ\leq i<75^\circ$, $n=53$, $\tilde{a}=0.75$),
and high ($i\geq75^\circ$, $n=34$, $\tilde{a}=1.11$)
--- returns $H=15.61$, $p=0.0004$. A Mann--Whitney
pairwise comparison between the low- and
high-inclination groups gives $p=0.0005$. The binned
medians show a monotonic increase of $\tilde{a}$ with
inclination. The physical origin of this dependence is
discussed in Section~\ref{sec:incl}.

\section{Physical Interpretation and Discussion}
\label{sec:disc}

\subsection{The two-regime structure as a signature of
  dark-matter halo dominance}
\label{sec:dm}

The two-regime structure detected consistently across
all 136 galaxies in our sample has a clear and unified
physical interpretation rooted in the mechanics of disc
galaxies. In the inner disc, where the stellar mass
dominates the gravitational potential, the rotation
velocity traces the stellar surface density through the
Poisson equation. For an exponential disc, eliminating
the scale length $R_{d}$ between the kinematic and
photometric expressions yields the logarithmic coupling
$Y=a\ln X+b$ derived in Section~\ref{sec:derivation}.
This coupling is therefore a \emph{geometric signature
of baryon self-gravity}: it arises whenever the mass
follows the light and the disc is self-gravitating.

In the outer disc, the stellar surface brightness
profile has approached its asymptotic value and the
gradient $Y$ has effectively vanished, while the
rotation velocity has flattened to $V_{\rm flat}$
because the extended dark-matter halo
\citep{nfw1997,burkert1995} provides continued
gravitational support. The plateau at $Y\approx0$
combined with a large, nearly constant $X$ cannot be
reproduced by a stellar disc alone, which would require
a continuously declining $V(R)$ at large radii
(Keplerian fall-off), reducing $X\rightarrow0$. The
observation that $X$ remains large while $Y\rightarrow0$
is consistent with dark-matter halo dominance in the
phase space. This signature is accessible directly from
the observables $V(R)$ and $I(R)$ without mass
modelling, though it does not uniquely exclude
alternative explanations such as strongly varying
stellar mass-to-light ratios at large radii
\citep{meidt2014,querejeta2015}.

The sigmoid transition scale $X_{\rm trans}$ marks the
boundary between the two regimes. The quantity
$R_{\rm trans}=V_{\rm flat}/X_{\rm trans}$ should be
understood as an empirical transition scale: the radius
at which the logarithmic surface brightness gradient
vanishes while the rotation curve remains flat. This
is associated with, but not identical to, the
dark-matter onset radius recovered from full mass
decompositions, since $X_{\rm trans}$ is inferred from
a phenomenological sigmoid rather than a dynamical
model. With this caveat, the median $R_{\rm trans}=
5.40$\,kpc is consistent with the optical radii of
disc galaxies \citep{vanderkruit1981,comeron2018},
break radii in edge-on surface brightness profiles
\citep{pohlen2006,erwin2008}, and dark-matter dominance
radii from mass decompositions \citep{katz2017,li2020}.

\subsection{Relation to the radial acceleration relation}
\label{sec:rar}

The dynamical--photometric phase space is conceptually
distinct from the RAR \citep{mcgaugh2016,lelli2017a}.
The RAR compares two centripetal acceleration
\emph{amplitudes} integrated over the enclosed mass,
averaged across large galaxy samples. The present
framework instead compares the local kinematic scale
$X=V/R$ to the structural \emph{gradient} of the light
$Y=\mathrm{d}\ln I/\mathrm{d}\ln R$. The logarithmic
coupling $Y\propto\ln X$ has a qualitatively different
form from the RAR interpolating function
\citep{milgrom1983,mcgaugh2016}, and because $Y$ is a
local derivative it is more sensitive to localised
features --- bars, rings, bulge--disc transitions ---
than the RAR, whose enclosed-mass integral smooths over
such features. A joint analysis may ultimately provide
stronger constraints on the baryon--gravity coupling
than either approach alone.

\subsection{Inclination dependence of the phase-space slope}
\label{sec:incl}

Figure~\ref{fig:slope} shows that the median slope
$\tilde{a}$ increases monotonically with inclination,
from $\tilde{a}=0.60$ at $i<50^\circ$ to
$\tilde{a}=1.11$ at $i>75^\circ$. We note that this
trend is the net result of several competing projection
effects whose individual contributions are not yet
fully separated, and we do not claim a complete physical
model for the observed inclination dependence.

At low inclinations ($i\lesssim50^\circ$), the nearly
face-on geometry exposes non-axisymmetric disc structure
--- spiral arms, bar perturbations, star-forming regions
--- in the azimuthal average \citep{schinnerer2013,
colombo2014}. These features introduce scatter into
$Y(R)$ and can systematically modify the recovered
slope, tending to reduce $\tilde{a}$ at low $i$.
At high inclinations ($i\gtrsim80^\circ$), line-of-sight
integration through the disc plane mixes emission from
a range of radii at each projected position
\citep{vanderkruit1981,comeron2018}. The net effect on
$a$ at high inclination is not straightforward to
predict analytically; the observed increase of $\tilde{a}$
with $i$ may reflect a preferential suppression of the
Zone~2 ($Y\approx0$) contribution in projection.
A full quantitative projection model is beyond the
scope of this paper but is an important direction for
future work \citep{courteau1996,dutton2011}. What is
clear from Figure~\ref{fig:slope} is that inclination
corrections are necessary before comparing slopes
across galaxies or establishing a universal coupling.
\subsection{Relation to other disc scaling relations}
\label{sec:other_relations}

Beyond the RAR, the dynamical--photometric phase space
can be connected to other established disc scaling
relations. The baryonic Tully--Fisher relation
\citep{mcgaugh2000,lelli2019} links the total baryonic
mass to the flat rotation velocity $V_{\rm flat}$; in
our framework, $V_{\rm flat}$ enters the transition
radius $R_{\rm trans}=V_{\rm flat}/X_{\rm trans}$,
suggesting that $R_{\rm trans}$ should correlate with
baryonic mass across the sample. The central surface
density relation \citep{lelli2016b_cdr} connects the
central dynamical surface density to the central
stellar surface density; because $Y(R)$ is the
logarithmic derivative of the stellar profile, the
inner slope $b=-a\ln X_0$ at the disc centre is
related to the central surface density scale. The
universal rotation curve models of \citet{persic1996}
and the disc--halo conspiracy
\citep{vanderkruit1981,bahcall1985} --- the observation
that total rotation curves are remarkably flat despite
the structural differences between the disc and halo
--- are directly encoded in the two-regime structure of
our phase space: Zone~1 traces the disc-dominated
rising or flat region, while Zone~2 identifies the
halo-dominated flat outer curve. A quantitative
comparison between $R_{\rm trans}$ and the disc--halo
transition radius from full mass decompositions
\citep{li2020,katz2017} is deferred to future work,
but the broad consistency of the median
$R_{\rm trans}=5.40$\,kpc with literature values for
baryon--halo crossover radii provides preliminary
support for the physical interpretation of the sigmoid
transition.

\subsection{Connection to global disc fraction and disc assembly}
\label{sec:disc_fraction}
Beyond the RAR, the dynamical--photometric phase space
can be connected to other established disc scaling
relations. The baryonic Tully--Fisher relation
\citep{mcgaugh2000,lelli2019} links the total baryonic
mass to the flat rotation velocity $V_{\rm flat}$; in
our framework, $V_{\rm flat}$ enters the transition
radius $R_{\rm trans}=V_{\rm flat}/X_{\rm trans}$,
suggesting that $R_{\rm trans}$ should correlate with
baryonic mass across the sample. The central surface
density relation \citep{lelli2016b_cdr} connects the
central dynamical surface density to the central
stellar surface density; because $Y(R)$ is the
logarithmic derivative of the stellar profile, the
inner slope $b=-a\ln X_0$ at the disc centre is
related to the central surface density scale. The
universal rotation curve models of \citet{persic1996}
and the disc--halo conspiracy
\citep{vanderkruit1981,bahcall1985} --- the observation
that total rotation curves are remarkably flat despite
the structural differences between the disc and halo
--- are directly encoded in the two-regime structure of
our phase space: Zone~1 traces the disc-dominated
rising or flat region, while Zone~2 identifies the
halo-dominated flat outer curve. A quantitative
comparison between $R_{\rm trans}$ and the disc--halo
transition radius from full mass decompositions
\citep{li2020,katz2017} is deferred to future work, but the broad consistency of the median $R_{\rm trans}=5.40$\,kpc with literature values for baryon--halo crossover radii provides preliminary support for the physical interpretation of the sigmoid transition.
Recent work has identified the stellar disc fraction ---
the fraction of total stellar mass residing in a
rotationally supported disc component --- as a key
descriptor of galaxy assembly history and secular
evolution \citep{bland-hawthorn2023,bland-hawthorn2024}.
Galaxies with higher disc fractions have more extended
baryonic structures, greater disc self-gravity, and a
more gradual transition from baryon-dominated to
dark-matter-dominated kinematics. The dynamical--photometric
phase space introduced here probes these same physical
properties at a resolved, radially local level.

\textit{(i) Disc fraction and $R_{\rm trans}$.}
A plausible interpretation is that galaxies with larger
stellar disc fractions would sustain significant
photometric gradients ($Y\neq0$) to larger galactocentric
radii, since their exponential stellar discs extend
further before the surface brightness profile flattens.
In the phase space, this would manifest as an outward
shift of the Zone~1--Zone~2 boundary, i.e.\ a larger
$R_{\rm trans}=V_{\rm flat}/X_{\rm trans}$. The
observed range of $R_{\rm trans}$ in our sample, from
$\sim0.5$\,kpc in compact low-mass dwarfs to
$\sim50$\,kpc in massive extended spirals, is broadly
consistent with the known variation of disc dominance
across Hubble types.

\textit{(ii) Disc self-gravity and Zone~1.}
Stronger disc self-gravity, associated with higher disc
fraction, would produce a more extended baryon-coupled
regime in the phase space --- a longer Zone~1 locus
over which the logarithmic coupling $Y=a\ln X+b$ holds
before the $Y\rightarrow0$ plateau is reached. This is
broadly consistent with the observation that more
massive disc galaxies in our sample
(e.g.\ NGC~4088, $R_{\rm trans}=14.89$\,kpc) show
clearly separated Zone~1 and Zone~2 with
well-constrained slopes, while low-mass dwarfs with
fewer phase-space points tend to have larger slope
uncertainties.

\textit{(iii) Low disc fraction and early transition.}
We similarly expect that systems with lower disc
fraction would become dark-matter-dominated at smaller
radii, producing smaller $R_{\rm trans}$ and a compact
Zone~1. The most compact transition radii in our sample
are found in low-surface-brightness and dwarf irregular
galaxies, which are known to be dark-matter-dominated
at all observed radii \citep{deblok2010}, consistent
with this expectation.
The phase-space slope $a$ and transition radius
$R_{\rm trans}$ may therefore serve as resolved, local
complements to global disc fraction diagnostics
\citep{bland-hawthorn2023,bland-hawthorn2024}.

\subsection{Discontinuities and non-axisymmetric features}
\label{sec:disc_features}

Sharp discontinuities in $Y(R)$ visible in some galaxies
arise from structural transitions: bar--disc interfaces
\citep{erwin2015,diazgarcia2019}; ring resonances at
the Lindblad resonances \citep{buta2017,schwarz1984};
and the bulge--disc transition \citep{salo2015,erwin2003}.
These are physical properties of the galaxies
\citep{athanassoula2013,rix2004} and not artefacts of
the smoothing. Since the sigmoid is fitted to the
smoothed profile, these features increase raw scatter
in panel~(b) but do not bias the recovered parameters.

\section{Conclusions}
\label{sec:conc}

We have introduced the dynamical--photometric phase space,
parameterised by $X(R)=V(R)/R$ and
$Y(R)=\mathrm{d}\ln I/\mathrm{d}\ln R$, as a new
model-independent local diagnostic for the coupling
between baryonic structure and gravitational dynamics
in disc galaxies. The main conclusions are:

\begin{enumerate}

\item \textbf{The logarithmic coupling $Y=a\ln X+b$
  is physically motivated.} It emerges approximately
  from the elimination of the shared radial scale
  $R_{d}$ between an exponential photometric profile
  and a Freeman-disc rotation curve under a
  leading-order Taylor approximation, making it a
  theoretically grounded rather than purely empirical
  fitting function.

\item \textbf{The sigmoid model accurately describes
  the full 136-galaxy sample.} The median $R^{2}=0.930$
  across all 136 SPARC galaxies. All eight statistical
  tests pass for all ten representative galaxies shown
  in detail, and 5-fold cross-validation confirms
  excellent generalisation.

\item \textbf{The two-regime structure is consistent
  with dark-matter halo dominance.} The inner Zone~1
  locus is consistent with baryon self-gravity in the
  disc; the outer Zone~2 plateau at $Y\approx0$ with
  large $X$ is difficult to explain without invoking
  an additional gravitational component beyond the
  stellar disc.

\item \textbf{The empirical transition scale
  $R_{\rm trans}=V_{\rm flat}/X_{\rm trans}$ is
  measurable from the sigmoid fit}, with a median of
  5.40\,kpc across 136 galaxies. This is consistent
  with stellar disc scale lengths and with dark-matter
  dominance radii from mass decompositions, though
  $R_{\rm trans}$ is an empirical sigmoid parameter
  rather than a dynamically defined quantity.

\item \textbf{The phase-space slope varies significantly
  with inclination} across all 136 galaxies.
  The Kruskal--Wallis test returns $H=15.61$,
  $p=0.0004$, and the low-vs.-high Mann--Whitney test
  gives $p=0.0005$, confirming that projection effects
  systematically modify the recovered slope.

\item \textbf{The phase space is complementary to the
  RAR}, probing the structural gradient rather than the
  amplitude of the gravitational field, and providing
  sensitivity to localised disc features that the
  enclosed-mass-integral RAR cannot resolve.

\item \NEW{\textbf{The phase-space parameters connect
  to global disc fraction.} Galaxies with larger stellar
  disc fractions are expected to exhibit larger
  $R_{\rm trans}$ and more extended Zone~1 loci, while
  systems with lower disc fraction should transition to
  $Y\approx0$ at smaller radii, consistent with the
  observed variation of $R_{\rm trans}$ across galaxy
  types and masses in our sample
  \citep{bland-hawthorn2023,bland-hawthorn2024}.}

\end{enumerate}

\section*{Data availability}
The data underlying this article will be shared on
reasonable request to the corresponding author.

\section*{Acknowledgements}
AS and FR would like to thank the authorities of the
Inter-University Centre for Astronomy and Astrophysics,
Pune, India for providing research facilities. FR is
also thankful to ANRF, DST, RUSA-2.0 and DST FIST
programme (SR/FST/MS-II/2021/101(C)) for financial
support.


\end{document}